\documentclass[submitting]{nst}

\usepackage{subfigure,dcolumn}
\usepackage[T2A,T1]{fontenc}
\usepackage[english]{babel}

\usepackage{listings}
\begin{document}

\title{The development of a high granular crystal calorimeter prototype of VLAST}
\thanks{Supported by the Scientific Instrument Developing Project of the Chinese Academy of Sciences (No. GJJSTD20210009), the Scientific Instrument Developing Project of the Chinese Academy of Sciences (No. PTYQ2024BJ0011), Special Support for Origins of Space Explorer (No. GJ11050103) and the National Natural Science Foundation of China (Grant No. 12022503, 12025504, 12125505, 12227805, 12273120).}

\author{Yan-shuo Zhang}
\affiliation{State Key Laboratory of Particle Detection and Electronics, University of Science and Technology of China, Hefei 230026, China}
\affiliation{Department of Modern Physics, University of Science and Technology of China, Hefei 230026, China}
\author{Qian Chen}
\affiliation{State Key Laboratory of Particle Detection and Electronics, University of Science and Technology of China, Hefei 230026, China}
\affiliation{Department of Modern Physics, University of Science and Technology of China, Hefei 230026, China}
\author{Deng-yi Chen}
\affiliation{Key Laboratory of Dark Matter and Space Astronomy, Purple Mountain Observatory, Chinese Academy of Sciences, Nanjing 210023, China}
\author{Jian-guo Liu}
\affiliation{State Key Laboratory of Particle Detection and Electronics, University of Science and Technology of China, Hefei 230026, China}
\affiliation{Department of Modern Physics, University of Science and Technology of China, Hefei 230026, China}
\author{Yi-ming Hu}
\affiliation{Key Laboratory of Dark Matter and Space Astronomy, Purple Mountain Observatory, Chinese Academy of Sciences, Nanjing 210023, China}
\author{Yun-long Zhang}
\email[Corresponding author, ]{ylzhang@ustc.edu.cn}
\affiliation{State Key Laboratory of Particle Detection and Electronics, University of Science and Technology of China, Hefei 230026, China}
\affiliation{Department of Modern Physics, University of Science and Technology of China, Hefei 230026, China}
\author{Yi-feng Wei}
\affiliation{State Key Laboratory of Particle Detection and Electronics, University of Science and Technology of China, Hefei 230026, China}
\affiliation{Department of Modern Physics, University of Science and Technology of China, Hefei 230026, China}
\author{Zhong-tao Shen}
\affiliation{State Key Laboratory of Particle Detection and Electronics, University of Science and Technology of China, Hefei 230026, China}
\affiliation{Department of Modern Physics, University of Science and Technology of China, Hefei 230026, China}
\author{Chang-qing Feng}
\affiliation{State Key Laboratory of Particle Detection and Electronics, University of Science and Technology of China, Hefei 230026, China}
\affiliation{Department of Modern Physics, University of Science and Technology of China, Hefei 230026, China}
\author{Jian-hua Guo}
\affiliation{Key Laboratory of Dark Matter and Space Astronomy, Purple Mountain Observatory, Chinese Academy of Sciences, Nanjing 210023, China}
\affiliation{School of Astronomy and Space Science, University of Science and Technology of China, Hefei 230026, China}
\author{Shu-bin Liu}
\email[Corresponding author, ]{liushb@ustc.edu.cn}
\affiliation{State Key Laboratory of Particle Detection and Electronics, University of Science and Technology of China, Hefei 230026, China}
\affiliation{Department of Modern Physics, University of Science and Technology of China, Hefei 230026, China}
\author{Guang-shun Huang}
\affiliation{State Key Laboratory of Particle Detection and Electronics, University of Science and Technology of China, Hefei 230026, China}
\affiliation{Department of Modern Physics, University of Science and Technology of China, Hefei 230026, China}
\author{Xiao-lian Wang}
\affiliation{State Key Laboratory of Particle Detection and Electronics, University of Science and Technology of China, Hefei 230026, China}
\affiliation{Department of Modern Physics, University of Science and Technology of China, Hefei 230026, China}
\author{Zi-zong Xu}
\affiliation{State Key Laboratory of Particle Detection and Electronics, University of Science and Technology of China, Hefei 230026, China}
\affiliation{Department of Modern Physics, University of Science and Technology of China, Hefei 230026, China}

\begin{abstract}
 Very Large Area gamma-ray Space Telescope (VLAST) is the next-generation flagship space observatory for high-energy gamma-ray detection proposed by China. The observation energy range covers from MeV to TeV and beyond, with acceptance of 10 m$\rm^2$sr. The calorimeter serves as a crucial subdetector of VLAST, responsible for high-precision energy measurement and electron/proton discrimination. This discrimination capability is essential for accurately identifying gamma-ray events among the background of charged particles. To accommodate such an extensive energy range, a high dynamic range readout scheme employing dual avalanche photodiodes (APDs) has been developed, achieving a remarkable dynamic range of 10$\rm^6$. Furthermore, a high granular prototype based on bismuth germanate (BGO) cubic scintillation crystals has been developed. This high granularity enables detailed imaging of the particle showers, improving both energy resolution and particle identification. The prototype's performance is evaluated through cosmic ray testing, providing valuable data for optimizing the final calorimeter design for VLAST.
\end{abstract}


\maketitle

\section{Introduction}\label{sec.I}
\nolinenumbers

Space-based gamma-ray astronomy offers unparalleled advantages for observing the universe. Unburdened by Earth's atmosphere, these observatories enjoy broad bandwidth coverage, enabling detection of gamma rays across a vast energy spectrum, from MeV to TeV. This wide coverage is crucial for studying diverse phenomena, from nuclear processes in stars to the extreme environments around black holes and pulsars. The continuous monitoring capability of space-based telescopes provides excellent temporal resolution, allowing scientists to track the evolution of transient events like gamma-ray bursts and flares from active galactic nuclei. Moreover, the stable platform of space minimizes background noise, leading to high measurement precision and improved sensitivity for detecting faint sources. These combined advantages make space-based gamma-ray detection a crucial tool for investigating fundamental questions in astrophysics and cosmology, including the nature of dark matter, the origin of cosmic rays, and the mechanisms driving powerful astrophysical phenomena ~\cite{bib:1, bib:2, bib:3, bib:4, bib:5}.

Several successful gamma-ray missions have been successfully carried out around the world. The Energetic Gamma-ray Experiment Telescope (EGRET) ~\cite{bib:6, bib:7} on the Compton Gamma-ray Observatory (CGRO) significantly advanced our understanding of high-energy gamma-ray sources. The Astro-rivelatore Gamma a Immagini Leggero (AGILE) ~\cite{bib:8} and the Fermi Large Area Telescope (Fermi-LAT) ~\cite{bib:9, bib:10, bib:11} have further expanded our knowledge, provided detailed maps of the gamma-ray sky and revealed a wealth of new sources ~\cite{bib:12, bib:13, bib:14, bib:15}. The Dark Matter Particle Explorer (DAMPE) ~\cite{bib:16, bib:17} focuses on precise measurements of high-energy cosmic rays ~\cite{bib:18, bib:19} and gamma rays ~\cite{bib:5, bib:20}, contributing to the search for dark matter signatures.

Building upon these achievements, the Very Large Area gamma-ray Space Telescope (VLAST) ~\cite{bib:21, bib:22, bib:23, bib:24, bib:25, bib:31} is proposed as the next-generation world-leading space-based gamma-ray observatory with significantly enhanced capabilities. With a sensitive area of approximately 10 m$\rm^2$, an order of magnitude larger than Fermi-LAT, VLAST will achieve unprecedented sensitivity across a wider energy range, enabling the detection of fainter sources and more detailed studies of known objects. Its superior energy resolution and precise particle track reconstruction capabilities will further enhance its scientific reach, allowing for more accurate measurements of energy spectra and search for potential dark matter signatures.

\begin{figure}[!htb]
     \centering
     \includegraphics[width=0.95\hsize]{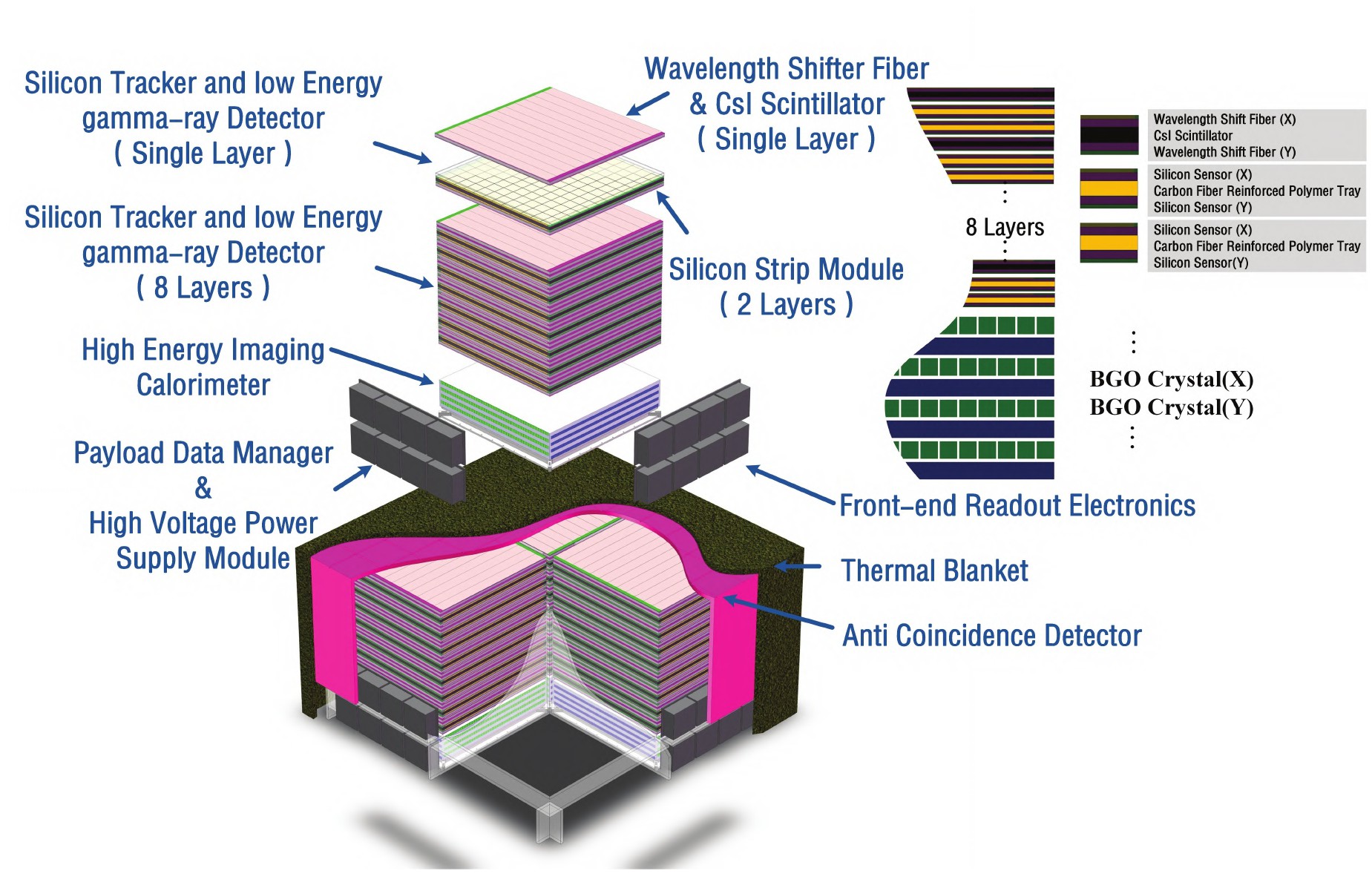}
     \caption{Architecture of the payload of VLAST.}
     \label{figure 1}
\end{figure} 

VLAST's payload instrument comprises three key components working in concert, as shown in Fig.~\ref{figure 1}. The outermost detector at the top, the Anti-Coincidence Detector (ACD) ~\cite{bib:23, bib:24}, is constructed from plastic scintillators and serves as a veto system to discriminate against charged particle background. The ACD also measures the energy loss of charged particles, aiding in the identification of light nuclei, while providing a trigger signal for inner detectors.

Beneath the ACD lies the Silicon Tracker and low Energy gamma Detector (STED) ~\cite{bib:25, bib:26}. Configured as a 2 × 2 array, each quadrant of the STED contains eight superlayers. Each superlayer integrates a thallium-doped cesium iodide (CsI(Tl)) detector and two double-sided silicon microstrip detectors. The CsI(Tl) detectors play a dual role: they directly measure the energy of MeV gamma rays and act as a converter for high-energy gamma rays, producing electron-positron pairs. The innovative usage of CsI(Tl) as a converter, instead of denser materials like tungsten, improves the energy measurement of lower-energy gamma rays. The silicon microstrip detector provides high angular resolution gamma-ray tracking and incident particle impact characterization through charged particle trajectory reconstruction.

At the base of the instrument is the High-Energy Imaging Calorimeter (HEIC), the heart of VLAST's high-energy measurements, which employs bismuth germanate (BGO) crystal as its primary sensitive material. The HEIC can accurately measure the characteristic profiles of the energy deposition of the secondary particles induced from electromagnetic shower and hadronic shower for efficiently distinguishing electrons (or gamma-ray) from hadrons. The calorimeter is designed to cover an extensive energy range from 0.1 GeV to 20 TeV for gamma photons and electrons, together with performance specifications of energy resolution better than 2 $\rm\%$ for 50 GeV photons, and the electron/proton separation capability should be better than 10$\rm^4$, which is essential for separating gamma-ray signals from the background of cosmic-ray protons.

The HEIC serves as a critical component in VLAST detectors, primarily responsible for the precise energy measurements and particle identification, so its design is a critical aspect of VLAST's development. Two potential design approaches are currently under consideration: (1) Following the DAMPE calorimeter design ~\cite{bib:27, bib:28, bib:29, bib:30}, which utilizes orthogonally arranged elongated BGO crystal bars ~\cite{bib:31}. While the long bar design offers potential advantages including better energy resolution, the manufacturing of 1.2-meter-long, high-quality BGO crystals presents significant technical challenges, and (2) Implementing a high-granular design composed of cubic BGO crystals, which offers a more practical approach, and its finer granular scheme enables more accurate electromagnetic shower profile measurement of energy deposition. The latter configuration shares many similarity with the CaloCube experiment ~\cite{bib:32, bib:33, bib:34, bib:35}, which was the first project consist of cubic crystals and has been validated to exhibit good performance.

This article presents the development and testing process of the cubic blocks scheme HEIC prototype, including its design, construction, and experimental validation, which are crucial steps towards finalizing the design of VLAST. The design of the prototype including the sensitive units, the readout electronics and the high dynamic range readout method are introduced in Section~\ref{sec.II}. Section~\ref{sec.III} details the construction of the HEIC-Cube prototype. In Section~\ref{sec.IV}, the cosmic ray test is conducted, validating the functionality of the prototype. Finally, Section~\ref{sec.V} concludes with a summary of key findings and implications.

\section{Design and development of HEIC-Cube prototype}\label{sec.II}

High granularity is a crucial design feature for modern calorimeters used in collider experiments and astro particle physics. It refers to the fine segmentation of the calorimeter into small, individually read-out sensitive elements. This fine segmentation allows for precise three-dimensional imaging of particle showers, enabling detailed reconstruction of the energy deposition pattern and improved particle identification.  This is particularly important for distinguishing between different types of particles, such as electrons, photons, and hadrons, and for reconstructing the complex topologies of high-energy particle interactions.

Several advanced calorimeter designs have been developed by the CALICE collaboration, showcasing different approaches to achieving high granularity. The silicon-tungsten (SiW) electromagnetic calorimeter utilizes silicon detectors interspersed with tungsten absorber plates ~\cite{bib:36}. Silicon detectors offer excellent spatial resolution, allowing for precise measurements of the shower development. The tungsten absorber provides the necessary material for the electromagnetic shower development. The scintillator-tungsten/copper (ScW) ECAL employs plastic scintillators and a tungsten-copper alloy absorber ~\cite{bib:37, bib:38}. Scintillators offer fast response times and good light yield, while the tungsten-copper absorber provides a compact design. The analog hadron calorimeter (AHCAL) is based on plastic scintillators and iron absorbers ~\cite{bib:39, bib:40, bib:41, bib:42}.  Iron is a cost-effective absorber material for hadronic calorimetry.  These sampling calorimeters, which alternate layers of active material (detector) and passive material (absorber), provide good imaging capabilities but often compromise with energy resolution due to the sampling fluctuations inherent in their design.

In contrast, homogeneous calorimeters, which are constructed entirely of active material, offer the potential for both superior imaging and better energy resolution. An example of this approach is the calorimeter design of the High Energy cosmic-Radiation Detection (HERD) experiment ~\cite{bib:43, bib:44, bib:45}, which utilizes an array of small-sized heavy crystals, LYSO, as sensitive units. The absence of passive absorber material in homogeneous calorimeters minimizes sampling fluctuations, leading to superior imaging performance and improved energy resolution. The HERD calorimeter, with its high granularity and homogeneous design, is optimized for measuring the energy and direction of high-energy cosmic rays.

The High Energy Imaging Calorimeter (HEIC) prototype described previously adopts a similar approach to HERD, using BGO crystals as the active material.  However, a key difference lies in the readout scheme. HERD employs two independent systems ~\cite{bib:46} to collect the scintillation light: the first one uses wavelength shifting fibers ~\cite{bib:47} to deliver the light to Intensified scientific CMOS (IsCMOS) cameras, whereas the second one makes use of photo-diode sensors ~\cite{bib:48, bib:49, bib:50}. Avalanche photodiodes (APDs) are selected for the HEIC prototype, and their associated electronic board are embedded into the calorimeter. The configuration shares a series of similarities with the HERD photodiode readout scheme, which has been validated in its previous experiment ~\cite{bib:51, bib:52}. This approach offers several advantages, including simplified integration of the readout electronics with the detector elements and reduced system complexity. This direct readout also reduces the potential for signal degradation associated with long optical fibers, and it improves light collection efficiency, which contributes to better energy resolution. The combination of high granularity, homogeneous crystal design, and fully embedded electronics makes the HEIC prototype a promising technology for future gamma-ray space observation.

We have developed a kind of high-granular BGO crystal calorimeter prototype designed for precise energy measurements of up to 50 GeV, and it is expected to exhibit a fine energy resolution for electrons and gamma-rays. To match the profile of the electromagnetic shower more accurately, the scalability in longitudinal depth is more significant than in horizontal direction. Therefore, the prototype consists of 10 longitudinal layers, each layer arranged in a 5 × 5 array of cubic BGO crystal with size of 3 cm × 3 cm × 3 cm. The calorimeter exhibits a total depth of approximately 27 radiation lengths, which is supposed sufficient for containing high-energy electromagnetic showers. The photoelectronic devices we chose are APDs, offering advantages in quantum efficiency and compactness when compared with photomultiplier tube (PMT) ~\cite{bib:53}. The electronic boards for signal processing are embedded inside the calorimeter structure, minimizing signal path lengths and optimizing performance. This design aims for a balance between granularity, energy resolution, and compactness, making it suitable for various high-energy physics experiments.

\subsection{Sensitive unit}

\begin{figure}[!htb]
\subfigure[]{
\label{figure 2a}
\includegraphics[width=0.305\hsize]{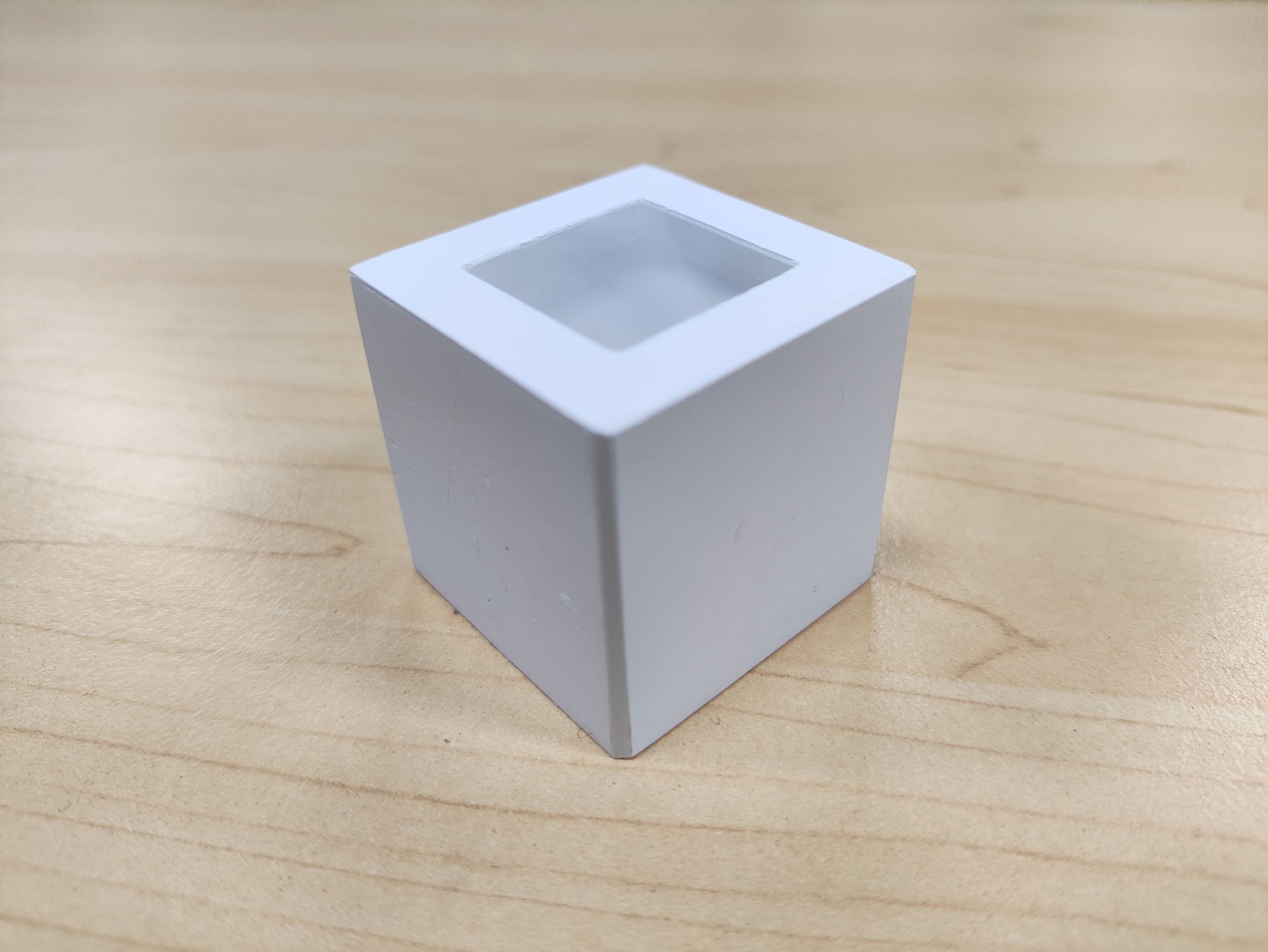}
}
\subfigure[]{
\label{figure 2b}
\includegraphics[width=0.23\hsize]{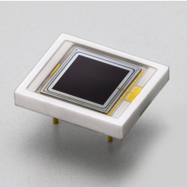}
}
\subfigure[]{
\label{figure 2c}
\includegraphics[width=0.37\hsize]{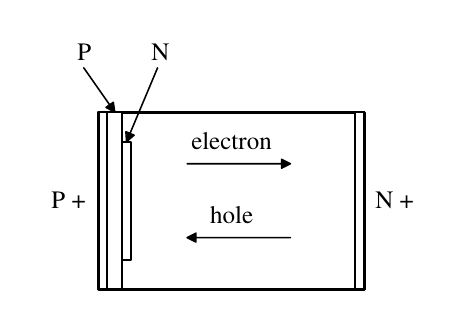}
}
\caption{(a) 30 mm side length BGO cubic crystal, coated with white reflective film. (b) The HAMAMATSU S8664-0505 type APD. (c) Vertical structure diagram of the APD.}
\label{figure 2}
\end{figure}

BGO crystals are widely applicable in high-energy physics experiments, including accelerator and space-based cosmic ray detection ~\cite{bib:54}, due to several key properties. Their high density allows efficient interaction with high-energy particles, while non-hygroscopicity ensures stable performance in varying environmental conditions. BGO's high light yield translates to better energy resolution, crucial for precise measurements. A group of inherent properties is listed in table~\ref{table 1}. Furthermore, the crystals are relatively easy to produce and process into desired shapes, simplifying detector construction. The HEIC-Cube prototype employs a set of BGO scintillators with dimensions of 30 mm × 30 mm × 30 mm, and each surface of the crystal is encased in a 0.3 mm thick barium sulfate (BaSO$\rm_4$) reflective coating, as depicted in Fig.~\ref{figure 2a}. This coating maximizes light collection by reflecting scintillation photons towards a designated 18 mm × 18 mm exit window on one surface of the cube, where a photodetector converts the light signal into an electrical pulse for subsequent analysis. This configuration optimizes light collection efficiency and relieves uneven light output to a certain extent, contributing to the detector's overall performance.

\begin{table}[!htb]
	\centering
	\caption{The properties of BGO [(Bi$\rm_2$O$\rm_3$)$\rm_2$(GeO$\rm_2$)$\rm_3$]}
	\label{table 1}    
	\begin{tabular}{ccc}
		\toprule
		Quantity & Value & Units \\
		\midrule
		Density & 7.130 & g cm$\rm^{-3}$ \\
		Radiation length & 1.118 & cm \\
		Molière radius & 2.259 & cm \\
		Minimum ionization & 8.918 & MeV cm$\rm^{-1}$ \\
		Nuclear interaction length & 22.32 & cm \\
		Fluorescence emission wavelength peak & 480 & nm \\
		Fluorescence decay time & 300 & ns \\
		\bottomrule
	\end{tabular}
\end{table}

When coupling with the emission wavelength of BGO scintillator, semiconductor photodetectors will have a better performance with higher quantum efficiency compared to PMTs ~\cite{bib:53}. The improved efficiency translates to a stronger signal for a given amount of scintillation light. After comprehensive consideration of critical parameters including device dimension, gain characteristic, and linearity range, the HAMAMATSU S8664 - 0505 avalanche photodiode (APD) ~\cite{bib:55} illustrated in Fig.~\ref{figure 2b} was selected for this study. The S8664-0505 features a 5 mm × 5 mm active area and employs a P-on-N structure ~\cite{bib:56}, as shown in Fig.~\ref{figure 2c}. The avalanche region P has an exceptionally thin profile of about 10 µm thick, which is crucial for minimizing unwanted signals from secondary particles. In this configuration, only the electrons in the P region generated through photoelectric conversion or ionization process can initiate the avalanche multiplication, amplifying the primary scintillation signal. In contrast, holes produced in the N- and N+ regions simply drift into the avalanche region without producing an avalanche. This unique structure effectively suppresses noise contributions from secondary particles generated by clustering effect directly within the N- and N+ regions of the APD. This targeted avalanche mechanism enhances the signal-to-noise ratio, improving the accuracy of energy measurements.

\begin{figure}[!htb]
     \centering
     \includegraphics[width=0.95\hsize]{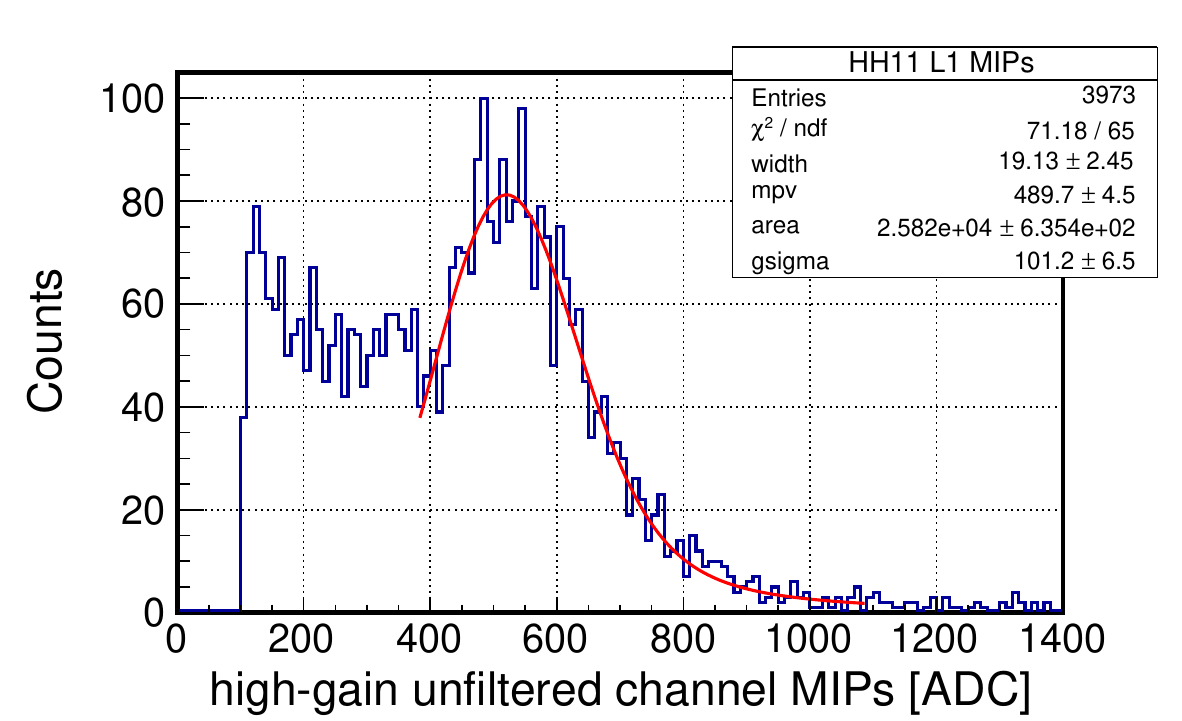}
     \caption{A typical MIPs distribution from a detector unit.}
     \label{figure 3}
\end{figure}

Considering the special environmental constraints of space-based experiments, the BGO crystal in our calorimeter prototype is intentionally not directly coupled with the APDs to mitigate device damage caused by mechanical vibrations. Instead, a 2 mm air (or vacuum) gap is maintained between the crystal surface and the APDs. This gap acts as a buffer, protecting the sensitive APDs from potential damage due to mechanical stresses and vibrations during launch and operation. The performance characteristics of this decoupled configuration was evaluated through measurements of its response to minimum ionizing particles (MIPs), as illustrated in Fig.~\ref{figure 3}. The signal amplitude is about 23.8 fC, corresponding to a light yield of roughly 2900 photoelectrons per MIP (pe/MIP). The maximum value of equivalent electronic noise in the high-gain channels is approximately 0.6 fC. This low noise level, combined with the substantial MIP signal, results in a favorable signal-to-noise ratio, demonstrating the effectiveness of this design even with the introduced air gap. This approach ensures the longevity and stability of the detector system in the demanding conditions of space.

\subsection{Electronics}

\begin{figure}[!htb]
     \centering
     \includegraphics[width=0.95\hsize]{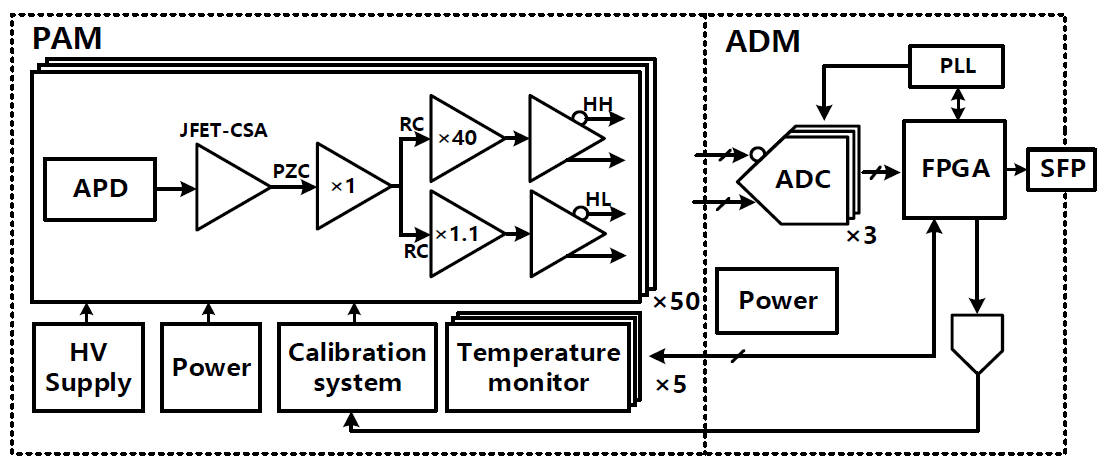}
     \caption{Block diagram of electronic pre-amplifier module (PAM) and analog to digital module (ADM).}
     \label{figure 4}
\end{figure}

The readout scheme adopts waveform sampling mode, capturing the complete signal shape from each APD. This system consists of two primary components: the Pre-Amplifier Module (PAM), and the Analog-to-Digital Module (ADM) ~\cite{bib:57, bib:58}, and they are interconnected via a FPGA Mezzanine Card (FMC) connector. This system performs pre-amplification and digitization of APD signals, as demonstrated in Fig.~\ref{figure 4}. Within the PAM, the JFETs are placed after each APD to suppress noise. A Pole-Zero Cancellation circuit (PZC) then shapes the Charge Sensitive Amplifier (CSA) output, which is subsequently split into high-gain and low-gain channels to extend the dynamic range.  This dual-gain approach allows for accurate measurement of both small signals from minimum ionizing particles and larger signals from high-energy depositions.

The ADM houses a 12-bit, 32-channel Analog-to-Digital Converter (ADC) operating at 40 MSPS, digitizing the differential signals from all gain channels, and then they are transmitted to the FPGA for storage in an internal buffer. Each individual waveform consists of 512 sampling points (corresponding to 12.8 µs), ensuring that both the baseline and the entire waveform are completely captured within the sampling window. The FPGA in ADM also serves as instruction parsing and clock distribution. Additionally, a temperature sensor is placed next to the central APDs to monitor its operational temperature for calibration and performance analysis. This circuit also includes a Digital-to-Analog Converter (DAC) calibration mechanism. The linearity of the high-gain and low-gain channels was obtained respectively from the DAC calibration, which will inject a group of step signals with varying amplitudes to the CSA unit. The results present that the response coefficients of the high and low gain channels are 20.6 ADC/fC and 0.53 ADC/fC, with corresponding dynamic ranges are 150 fC and 7000 fC, respectively.

The digitized signals are transmitted to a data acquisition board through optical fiber. The Data Concentrator Module (DCM) ~\cite{bib:59} primarily aggregates and buffers waveform data from multiple ADMs before uploading the consolidated data to the host computer via Ethernet for offline analysis and storage. Furthermore, the DCM incorporates essential functionalities including hit analysis, trigger generation, and command issuance. This hierarchical data acquisition system, from individual APDs to the central data collection, is designed for efficient and robust data handling in the laboratory environment. The inclusion of online data processing capabilities further optimizes data flow and reduces the volume of data transmitted to the back-end host.

\subsection{High dynamic range readout}

The VLAST calorimeter is designed to detect primary cosmic rays with energies exceeding 20 TeV. Simulation results indicate that for electrons/photons of primary energy of 20 TeV, the deposited energy approaches 7 TeV in a single crystal unit located in the shower center. The maximum energy in one crystal corresponds to approximately 2.4 × 10$\rm^5$ MIPs, where MIP represents the energy deposition by a minimum ionization particle through a BGO crystal of 30 mm thickness, which is approximately 27.4 MeV based on simulations. To ensure accurate reconstruction of high-energy particle showers as well as reduce false triggering from electronic noise, the energy lower threshold is set to 0.1 MIPs for each detection unit. This low threshold is essential for capturing the full extent of the particle showers and distinguishing them from background noise. This condition necessitates the readout scheme of the calorimeter to achieve a dynamic range of 10$\rm^6$.

If a single readout device is placed under a BGO crystal for fluorescence collection, its dynamic range is limited by random noise and saturation effect, so it cannot satisfy the design specification of the large area calorimeter. To overcome these limitations, many international studies have developed photodetectors or readout electronics with multiple gain stages by providing progressive sensitivity levels to extend the dynamic range. For example, the calorimeter of Fermi-LAT adopts a pair of photodiodes with distinct sensitive areas ~\cite{bib:9, bib:10} coupled to each end of individual scintillator. Furthermore, each of their readout electronics incorporate dual-gain channels to achieve additional dynamic range expansion, which ensures the effective energy range coverage from 20 MeV to about 300 GeV.

Following the adoption route of semiconductor photodetectors, the HEIC employs a dual-APD configuration, where two APDs are positioned adjacent to each other in accordance with the scintillator geometry and coupled to the same fluorescence exit window of every BGO cubic units. These APDs are arranged directly facing the crystal's light-emitting surface to ensure optimal fluorescence collection efficiency. A crucial element of the HEIC design is the incorporation of a light intensity attenuation filter. The entrance windows of one APD between the pair are covered with a fluorescence attenuation filter for extending the dynamic ranges for the high-energy end. The attenuation filter, an exposure film, is made of a plastic sheet with a thickness of 0.17 mm. It is designed to tune the amplitudes of signals by proportional attenuating the intensity of scintillation light injected into the APDs, which remains it within the linear operating range of the electronic amplifiers. The readout electronics used for both APDs are identical, and each APD leads out both an initial channel and an amplified channel on the back of the front-end board. This dual-APD configuration, combined with dual-gain channels in the readout electronics for each APD, provides the necessary dynamic range.

\begin{figure}[!htb]
     \centering
     \includegraphics[width=0.95\hsize]{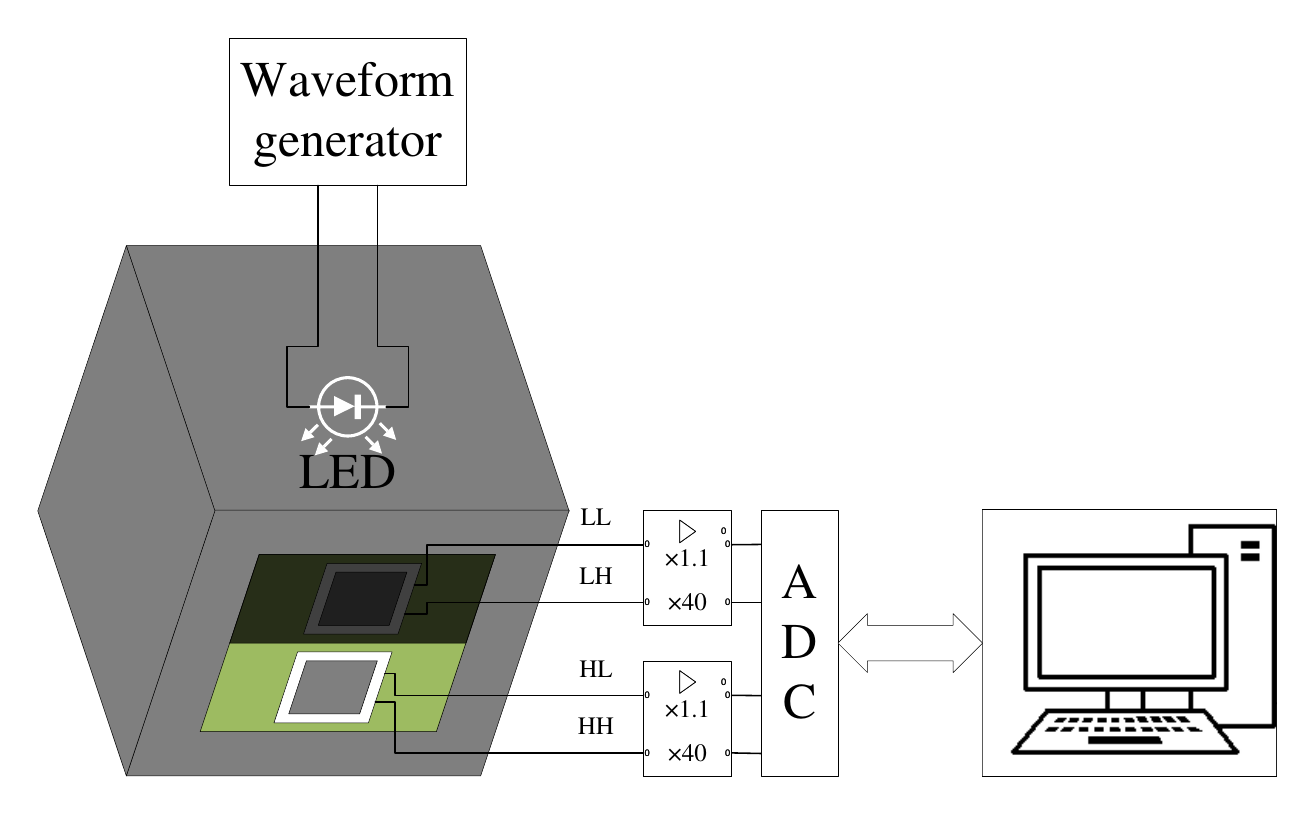}
     \caption{Block diagram of the devices that utilizes a LED for measuring gain ratio.}
     \label{figure 5}
\end{figure}

To validate this large dynamic range readout scheme, an LED-based illumination system was established, as depicted in Fig.~\ref{figure 5}. The readout system was enclosed in a darkroom environment to minimize stray light interference, with a LED positioned above the APD array. The LED served as a light source, simulating the scintillation light produced by the BGO crystal when interacting with cosmic rays. The pulse generator was used to excite LED, where the light intensity was precisely controlled through pulse voltage modulation ~\cite{bib:60}. The pulse duration was set to 300 ns which is consistent with the fluorescence decay time of BGO crystal. The signals from two APDs were independently processed by the readout electronics, with one APD covered by a filter with an attenuation coefficient about 1000 times placed on the incident window. Each APD signal is processed through two channels: a high-gain and a low-gain channel. This results in four readout channels per crystal: HH (High-gain from the unfiltered APD), HL (Low-gain from the unfiltered APD), LH (High-gain from the filtered APD), and LL (Low-gain from the filtered APD).

\begin{figure}[!htb]
\subfigure[]{
\label{figure 6a}
\includegraphics[width=0.3\hsize]{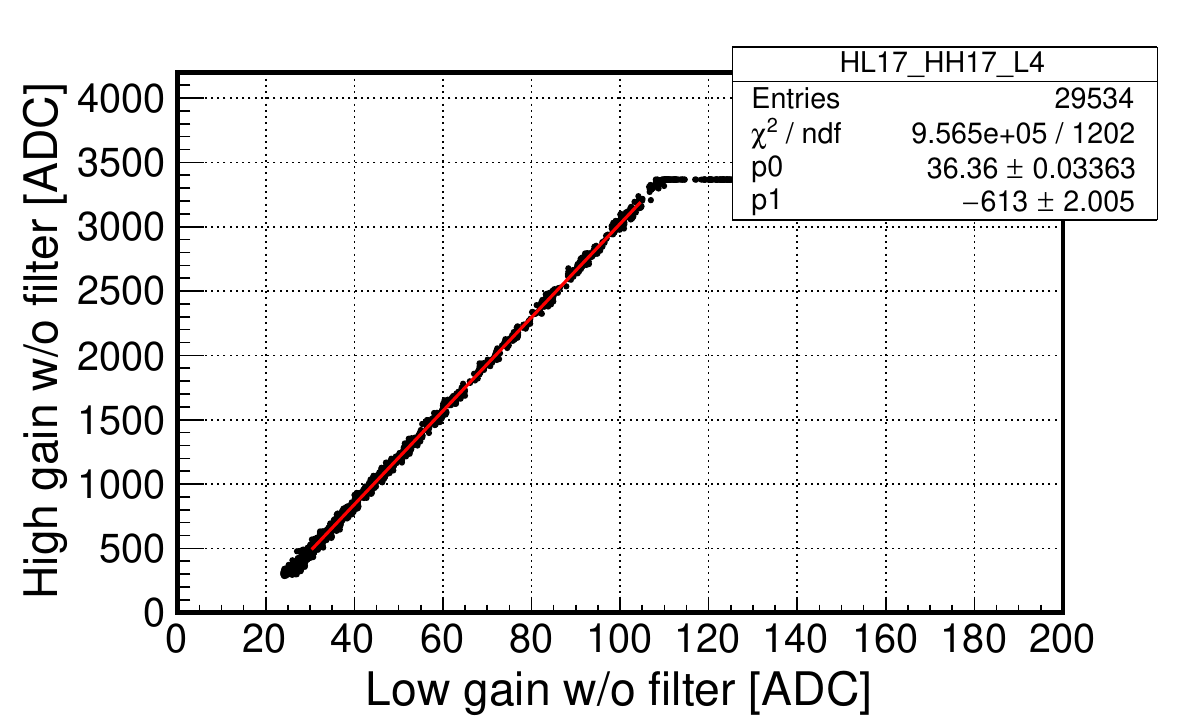}
}
\subfigure[]{
\label{figure 6b}
\includegraphics[width=0.3\hsize]{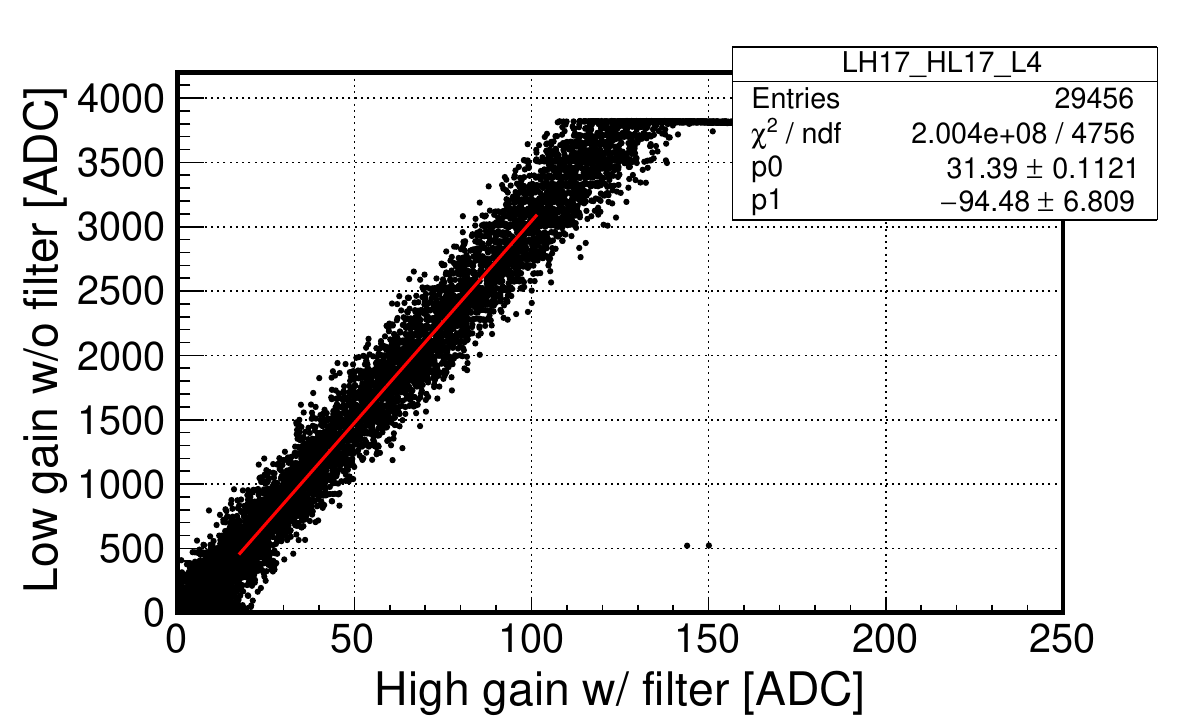}
}
\subfigure[]{
\label{figure 6c}
\includegraphics[width=0.3\hsize]{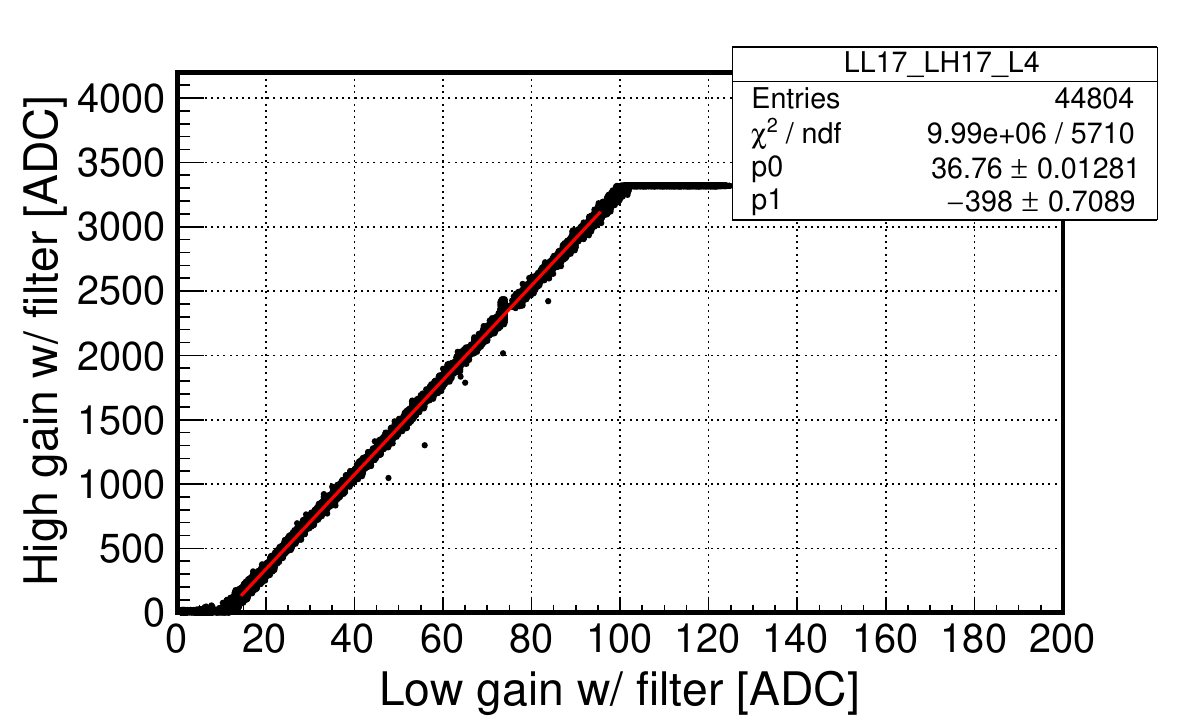}
}
\caption{The linear fitting results between different channels. (a) from the uncovered APD: high-gain vs low-gain; (b) channels between 2 APDs: low-gain without filter vs high-gain with filter; (c) from the covered APD: high-gain vs low-gain.}
\label{figure 6}
\end{figure}

During the testing process, the voltage of the pulse generator was continuously adjusted to modulate the LED light intensity, which allows for a systematic investigation of the readout system's response across a wide range of simulated energy depositions. The gain ratios between adjacent readout channels were carefully measured and analyzed, as presented in Fig.~\ref{figure 6}. Linear fitting analysis revealed that the HH-HL and LH-LL ratios both exhibited values of approximately 36.5, which is determined by the design of the electronics. Figure~\ref{figure 6a} and ~\ref{figure 6c} depict their amplitude correlations between high-gain and low-gain channels respectively. The HL-LH ratio was measured at about 31, a parameter introduced by the attenuation filter, illustrated in Fig.~\ref{figure 6b}. Based on these ratios and the MIPs response characteristics of the BGO sensitive unit from Fig.~\ref{figure 3} (showing a signal amplitude of 23.8 fC for a single MIP), the effective energy range of the four channel responses is detailed in Table~\ref{table 2}. The combined dynamic range will exceed 2.4 × 10$\rm^6$ according to the relative gains between the readout channels, which proves to fulfill the design specifications for large-area calorimeter applications, demonstrating the effectiveness of the dual-APD, dual-gain readout scheme.

\begin{table}[!htb]
    \centering
    \caption{The electronic and system dynamic range of different readout channels}
    \label{table 2}    
    \begin{tabular}{ccc}
    \toprule
    Readout channel & Electronic dynamic range & System dynamic range \\
    \midrule
    HH & 3 fC – 150 fC & 0.13 MIPs – 6.30 MIPs \\
    HL & 10 fC – 7 pC  & 0.42 MIPs – 294 MIPs \\
    LH & 3 fC – 150 fC & 126 MIPs – 6300 MIPs \\
    LL & 10 fC – 7 pC  & 420 MIPs – 294000 MIPs \\
    \bottomrule
    \end{tabular}
\end{table}

However, the LED-based illumination system has limitations. The light emitted by the LED is not uniform in all directions, and may not be evenly distributed across the APD array, leading to variations in the measured signals. This can affect the accuracy of the gain ratio measurements and the determination of the dynamic range. Therefore, further experimental validation using more uniform and representative radiation sources, such as radioactive sources or particle beams, is crucial for precise calibration and performance verification of the dual-APD, dual-gain readout system. This additional validation will provide a more accurate assessment of the system's performance and its ability to accurately measure the energy of cosmic rays over the targeted energy range.

\section{Construction of the HEIC-Cube prototype}\label{sec.III}

\begin{figure}[!htb]
     \centering
     \includegraphics[width=0.95\hsize]{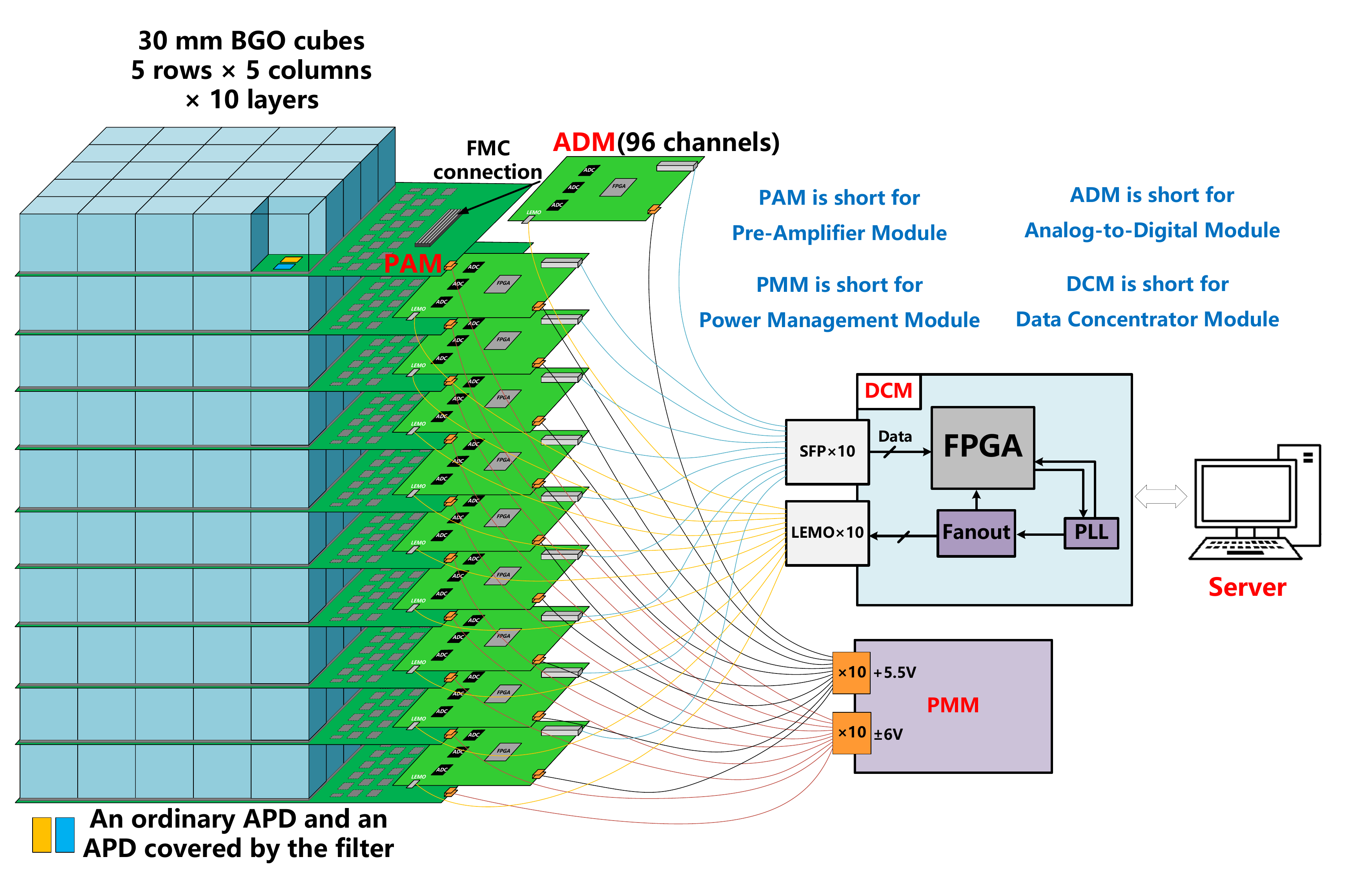}
     \caption{The schematic plot of the high granular crystal calorimeter prototype.}
     \label{figure 7}
\end{figure}

To validate the calorimeter performance in experimental conditions, a small-scale prototype was developed and tested in the laboratory. This prototype allows for a detailed investigation of the calorimeter's capabilities and serves as a crucial step towards the development of the full-scale instrument. Fig.~\ref{figure 7} illustrates a schematic diagram of its structure configuration. The prototype has 10 layers in vertical to ensure minimal leakage at the tail of the electromagnetic shower, allowing for a more accurate measurement of the total deposited energy. Each layer contains a 5 × 5 array of BGO crystals with the size of 3 cm × 3 cm × 3 cm, yielding a sensitive area of 15 cm × 15 cm per layer, sufficient to cover the lateral spread of a shower. A PAM board responsible for amplifying the signals from the APDs is embedded at the base of each sampling layer, and it is interconnected with an ADM board via a FMC connector. The ADM board performs the analog-to-digital conversion of the amplified signals. Each ADM employs 3 ADC chips, providing a total of 96 effective input channels for digitization. Given the matching number of channels in PAM and ADM, signals from the four vertex channels (LL0, LL4, LL20, and LL24) exhibiting the smallest amplitudes in each layer are excluded and remain unconnected to the back-end board. These channels corresponding to the corners of the crystal array are expected to contribute less significantly to the overall energy measurement.

\begin{figure}[!htb]
\subfigure[]{
\label{figure 8a}
\includegraphics[width=0.45\hsize]{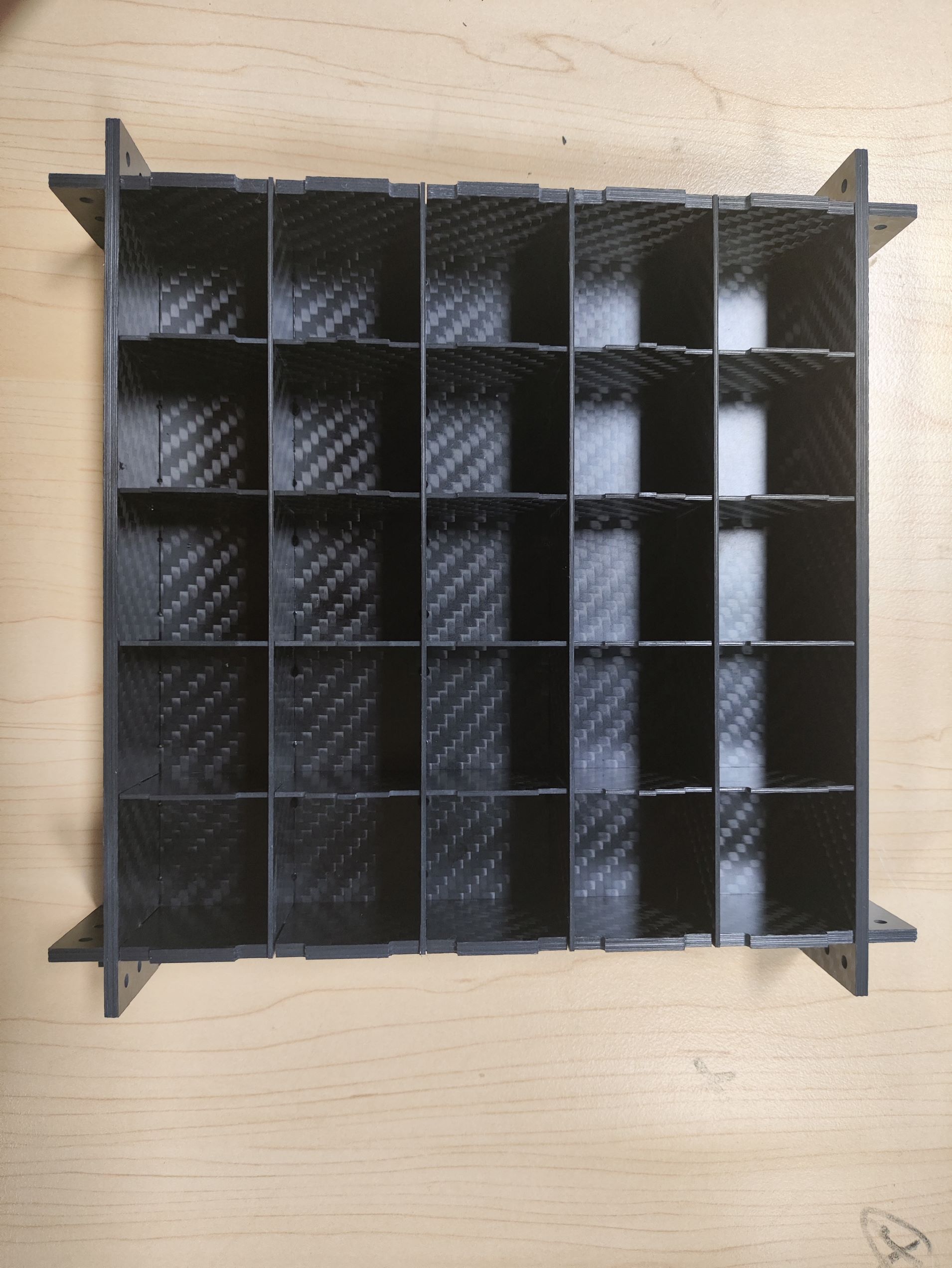}
}
\subfigure[]{
\label{figure 8b}
\includegraphics[width=0.45\hsize]{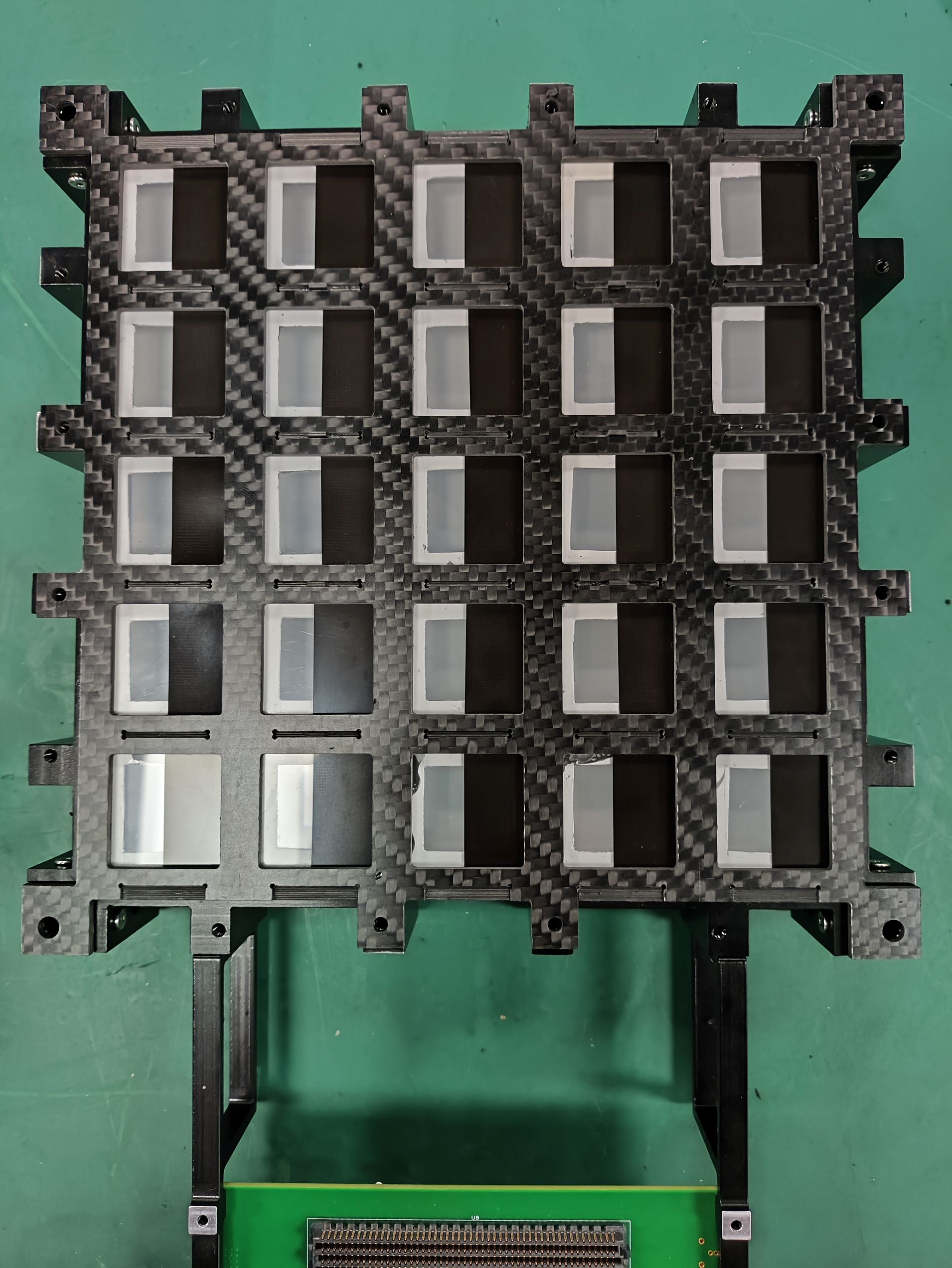}
}
\subfigure[]{
\label{figure 8c}
\includegraphics[width=0.45\hsize]{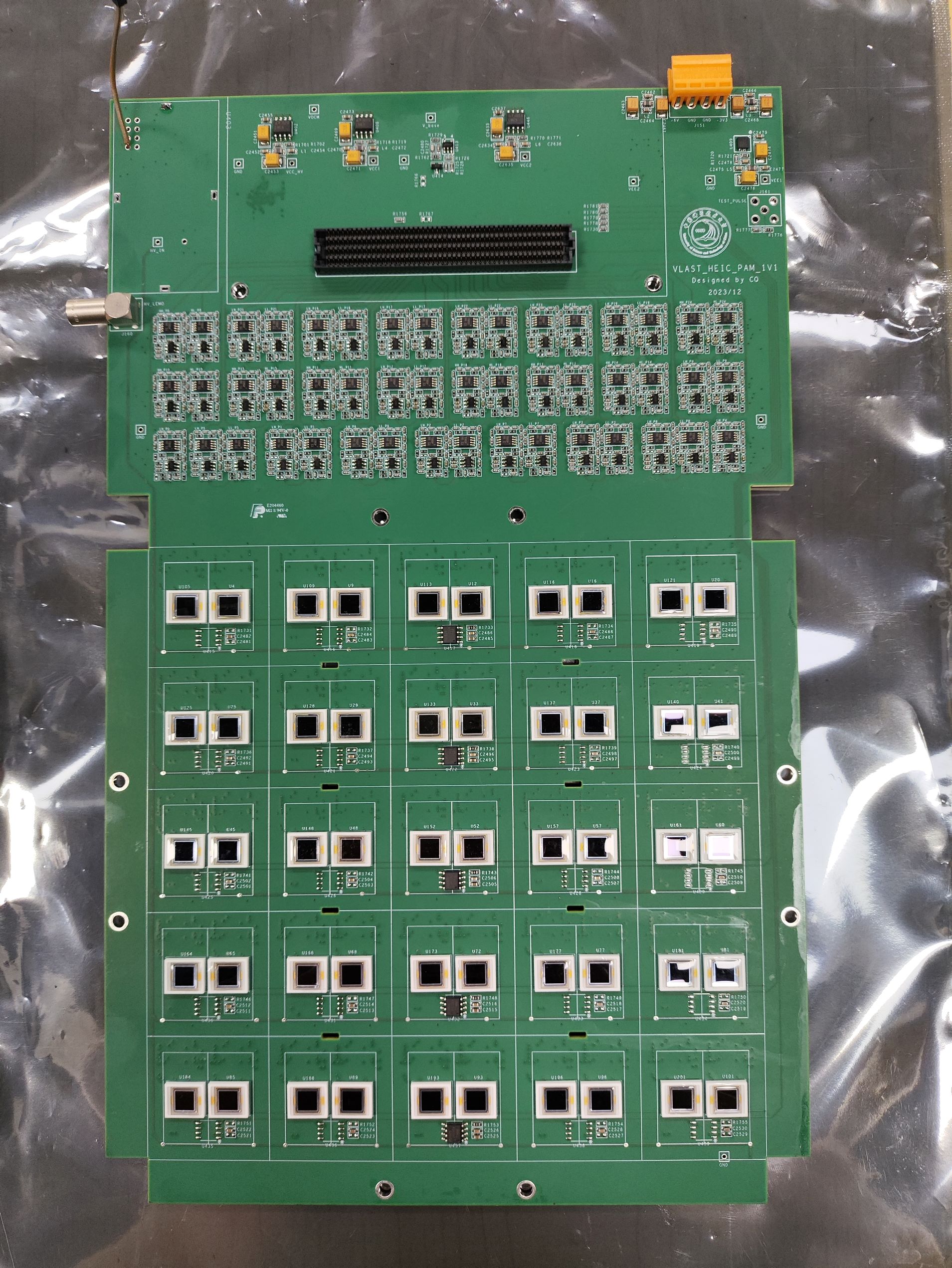}
}
\subfigure[]{
\label{figure 8d}
\includegraphics[width=0.45\hsize]{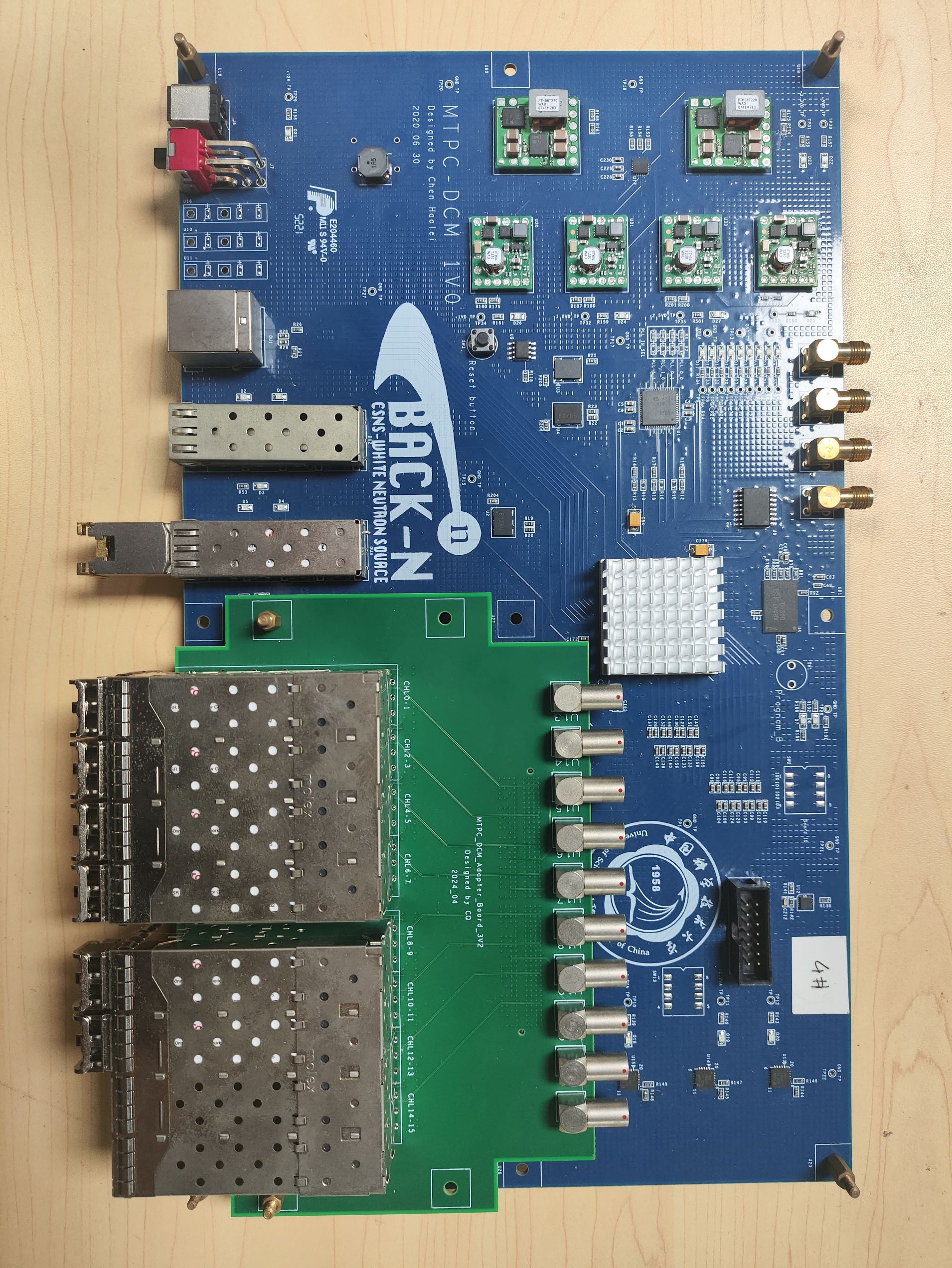}
}
\subfigure[]{
\label{figure 8e}
\includegraphics[width=0.95\hsize]{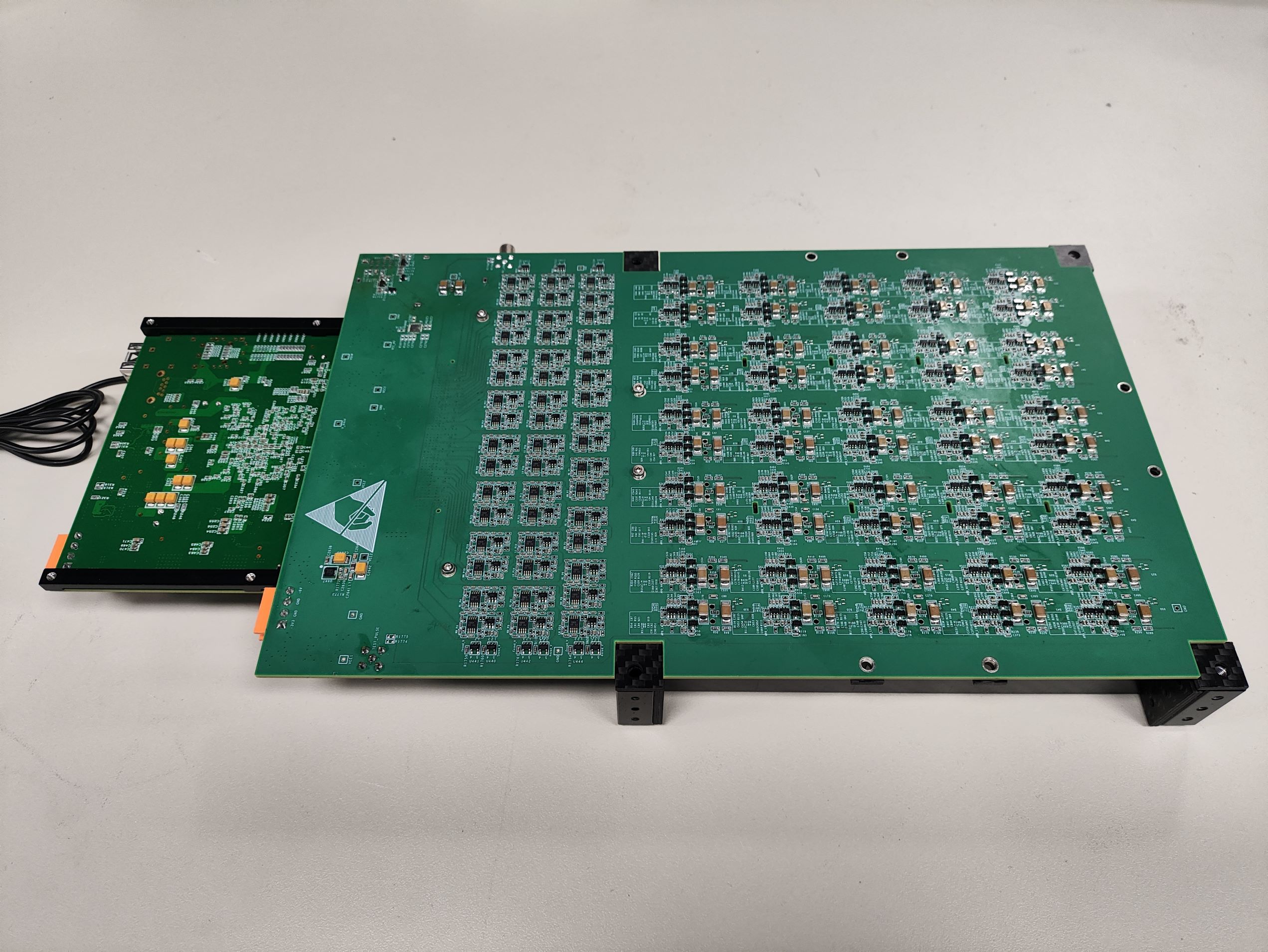}
}
\caption{(a) The carbon fiber framework. Each grid will house a BGO block. (b) The framework assembled with BGO cubes and plastic attenuators. (c) the end-product of PAM. It is interlocked with the peripheral carbon fiber framework of a BGO crystal layer. (d) the back-end DCM. (e) an entire overview of the calorimeter prototype component with a BGO crystal layer at the bottom.}
\label{figure 8}
\end{figure}

Prior to the construction of the calorimeter, a pairing process is carried out based on the fluorescence output of each BGO crystal and the gains of its corresponding APDs. This ensures a better consistency in the response of each detection unit to MIPs. These matched crystal-APDs pairs are then installed in corresponding positions within the same layer to maintain similar performance characteristics. The mechanical design of the prototype is carefully considered to ensure the precise positioning and stability of the BGO crystals and readout electronics. Fig.~\ref{figure 8a} shows the box of the sensitive layer made of carbon fiber, a lightweight and strong material, which consists of a 5 × 5 array of arranged cells. Each cell measures 31.6 mm × 31.6 mm to accommodate a BGO crystal placement with a small tolerance for positioning and thermal expansion. The intercellular partitions have a uniform thickness of 1 mm, while the outermost frames exhibit a 2 mm thickness for additional structural reinforcement. Fig.~\ref{figure 8b} demonstrates the assembled configuration, where BGO crystals are precisely positioned within their respective cell and covered with attenuation filters above their optical windows. The filters are crucial for the dynamic range extension of the readout system, as they attenuate the light reaching one of the two APDs coupled to each crystal, which allows the system to measure a wider range of energies without saturation. The crystals are securely fixed in place using a black shock-absorbing adhesive (DOWSIL SE 9186 L Black Sealant). This adhesive provides both mechanical stability and optical isolation. Additionally, a carbon fiber honeycomb structure is placed above the filters to maintain the required air gap.

The prototype readout electronic board is presented in Fig.~\ref{figure 8c}, housing the readout for 25 crystals arranged in a horizontal layer. APDs have been soldered onto the PCB for efficient signal transfer and mechanical stability, and each crystal unit interfaces with 2 APDs. They share the same voltage supplied by a Nuclear Instrumentation Module (NIM) power source, and a compact high voltage low dropout regulator (LDO) will be installed at the top right-hand corner of PAM during subsequent assembly. The consistency of the high-gain and low-gain coefficients in the readout electronics is relatively straightforward to control. The PAM board is securely buckled onto the carbon fiber box according to locating bolts, while the corresponding ADM modules are connected and firmly fixed using screws, thereby constituting a complete sensitive layer as depicted in Fig.~\ref{figure 8e}. The complete prototype configuration integrates 10 identical layers in vertical stacking and a corresponding back-end DCM (Fig.~\ref{figure 8d}). Following assembly, each detection unit must still undergo calibration using MIPs. Their responses are subsequently normalized based on the respective MIP signal amplitude, thereby harmonizing the output across all units and enhancing the overall uniformity and measurement reliability of the detector system.

In summary, the high granular crystal calorimeter prototype comprises 10 detection layers, with each layer consisting of 25 BGO crystals with a side length of 30 mm. Each crystal is coupled with two APDs, generating 4 distinct amplitude output signals. This design results in a system-wide configuration of 250 crystals and 960 active readout channels. Accordingly, the detector's electromagnetic shower containment capability is characterized by 3.3 Molière radii in the transverse dimension and 26.8 radiation lengths in the longitudinal dimension when particles hit the calorimeter along the central axis. These parameters ensure that the prototype is well-suited for precise electron energy measurement in the GeV range. The prototype provides valuable experimental data for validating the performance of the dual-APD, dual-gain readout scheme and for characterizing the calorimeter's response to electromagnetic showers, paving the way for the development of the full-scale VLAST calorimeter.

\section{Performance of the prototype system}\label{sec.IV}

\begin{figure}[!htb]
     \centering
     \includegraphics[width=0.95\hsize]{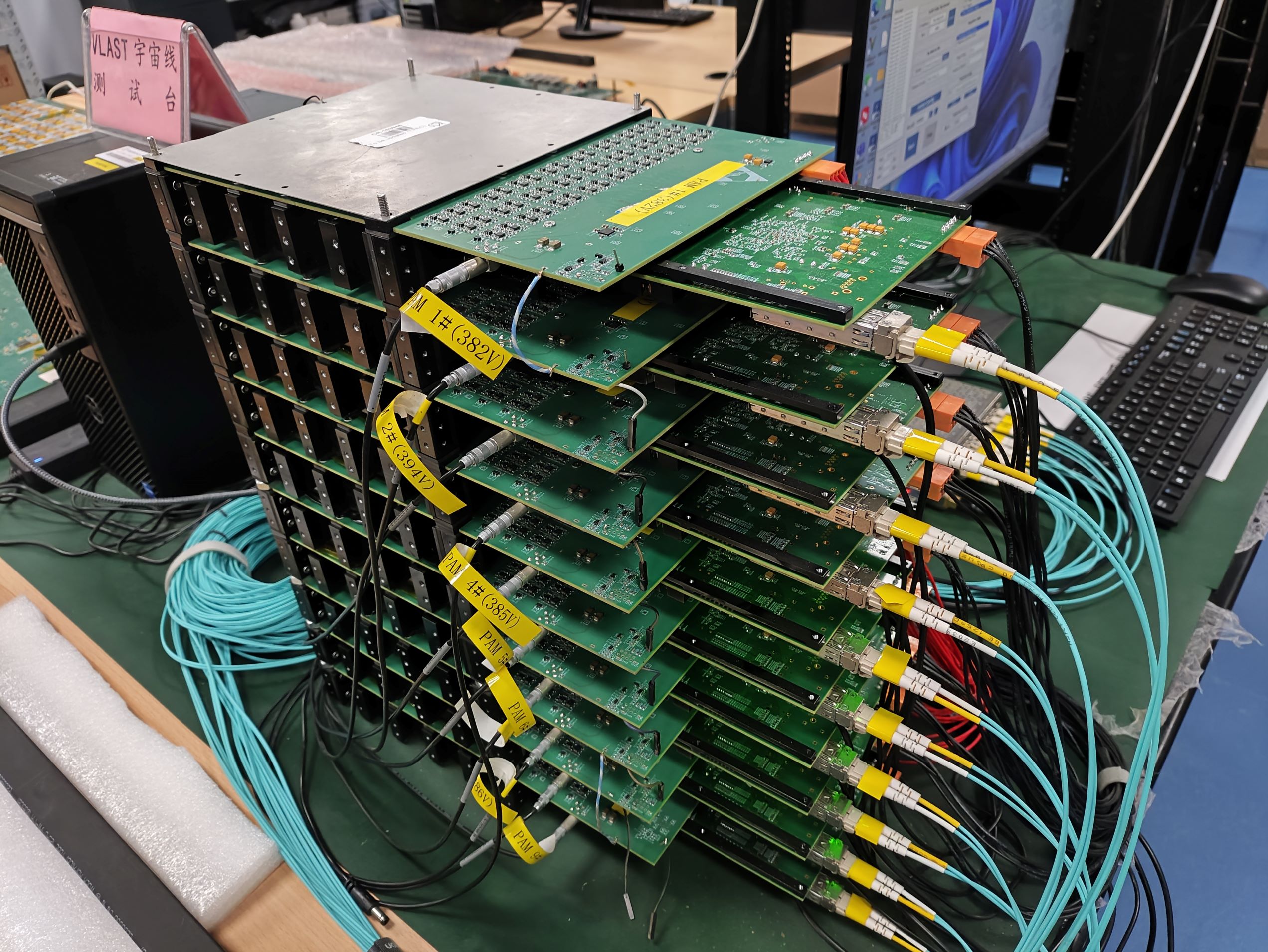}
     \caption{Overview photograph of the calorimeter prototype when conducting cosmic ray test.}
     \label{figure 9}
\end{figure} 

To further validate the calorimeter prototype and assess its performance under real-world conditions, an experimental setup was established for ground-based cosmic ray measurements. This testing involves exposing the prototype to the natural flux of cosmic rays, which serve as a readily available source of high-energy particles, providing an opportunity to evaluate its response. Fig.~\ref{figure 9} demonstrates the experimental setup for this cosmic ray test.

Muons, being highly penetrating particles, are the dominant component of cosmic rays reaching the Earth's surface. They readily traverse the calorimeter, depositing energy primarily through ionization, typically interacting with the calorimeter as MIPs. This behavior makes them ideal for characterizing the performance consistency among the individual detector units within the calorimeter. By analyzing the signals generated by muons passing through the detector, it's possible to identify any variations in response between different channels and assess the overall uniformity of the calorimeter.

The data acquisition system employed in this experiment utilizes an over-threshold triggering mode, which means that data is recorded only when the signal in a readout channel exceeds a predefined threshold. This threshold is set to 20 times the sigma value relative to its respective pedestal average for each channel. The pedestal represents the baseline signal level of each channel when no particle is interacting with the detector, while its width reflects the level of noise. Setting the threshold at 20 times the sigma value (approximate 130 ADC values, equivalent to 0.3 MIP MPV) helps to discriminate real signals from electronic noise fluctuations.

The experiment was conducted over a period of two months, accumulating a substantial amount of data. After excluding data associated with mistaken operations or other anomalies, a total of 59 valid data packages were obtained. The daily binary recording files indicate an interaction rate of approximately one million events per day. This high event rate provides a statistically significant dataset for analysis.

Some preliminary analysis results are presented here from the routine cosmic ray test. The offline data processing involves several steps to extract meaningful information from the raw data. First, the binary files are unpacked to retrieve the 512 sampling points datasets for each channel. These sampling points represent the digitized signal waveform over time. The average value of the first 128 sampling points is calculated and used as the pedestal value for that channel. The maximum code value among all 512 points is identified as the channel's peak value, representing the maximum amplitude of the signal. The effective amplitude of each channel is then determined by subtracting the pedestal value from the peak value. This procedure is applied consistently across all daily data packets to ensure uniformity in the analysis.

\begin{figure}[!htb]
\subfigure[]{
\label{figure 10a}
\includegraphics[width=0.45\hsize]{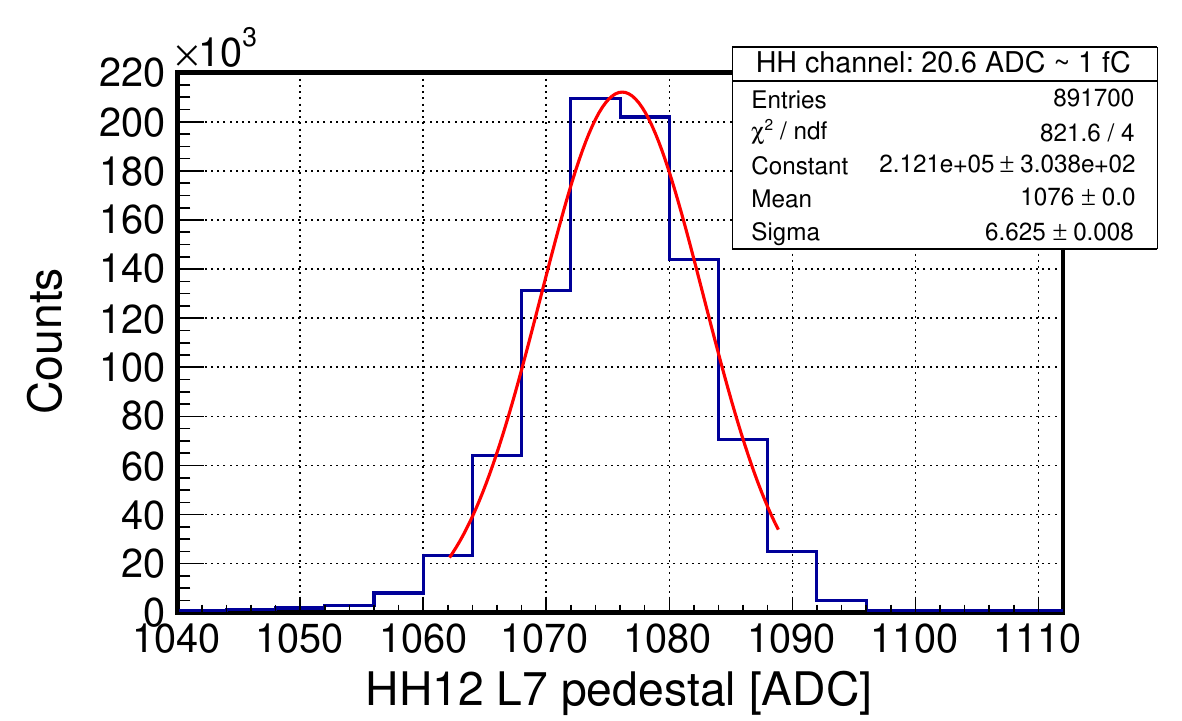}
}
\subfigure[]{
\label{figure 10b}
\includegraphics[width=0.45\hsize]{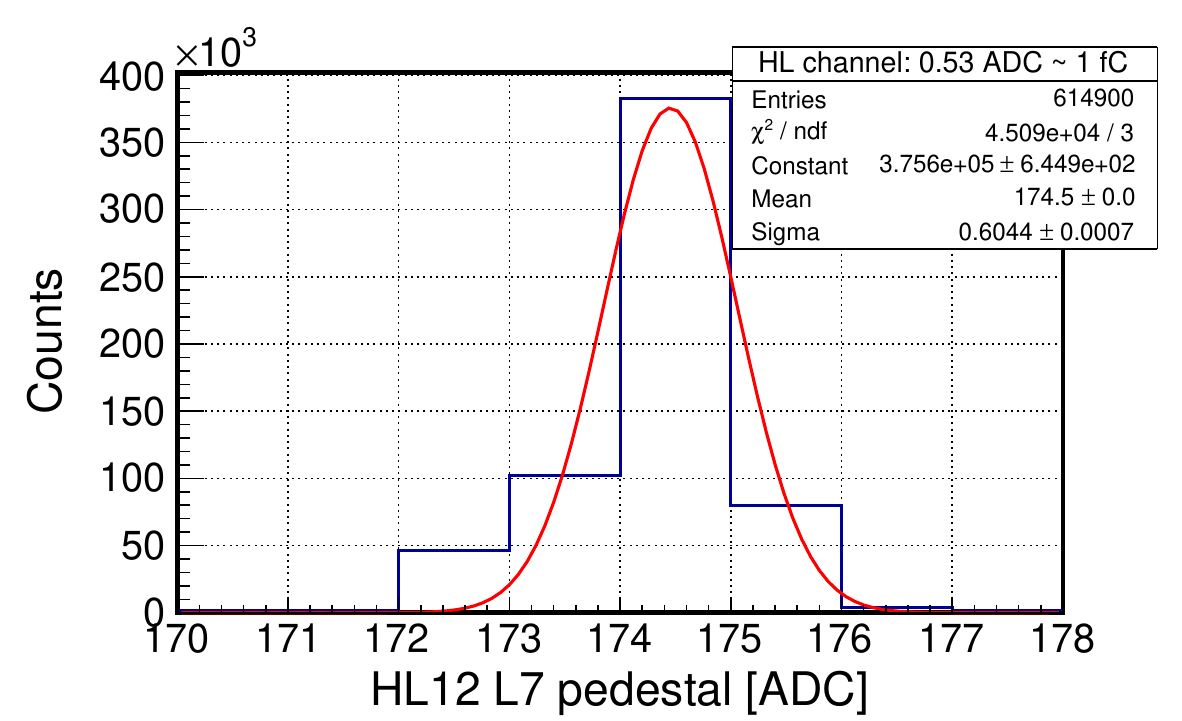}
}
\subfigure[]{
\label{figure 10c}
\includegraphics[width=0.45\hsize]{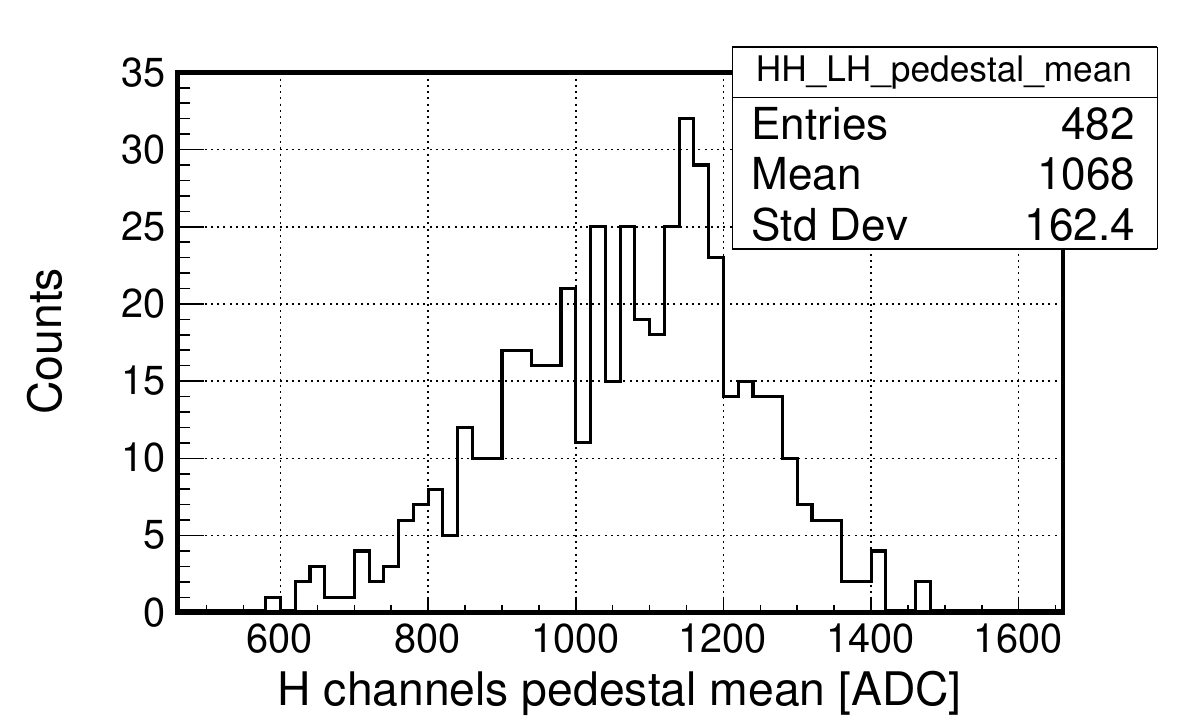}
}
\subfigure[]{
\label{figure 10d}
\includegraphics[width=0.45\hsize]{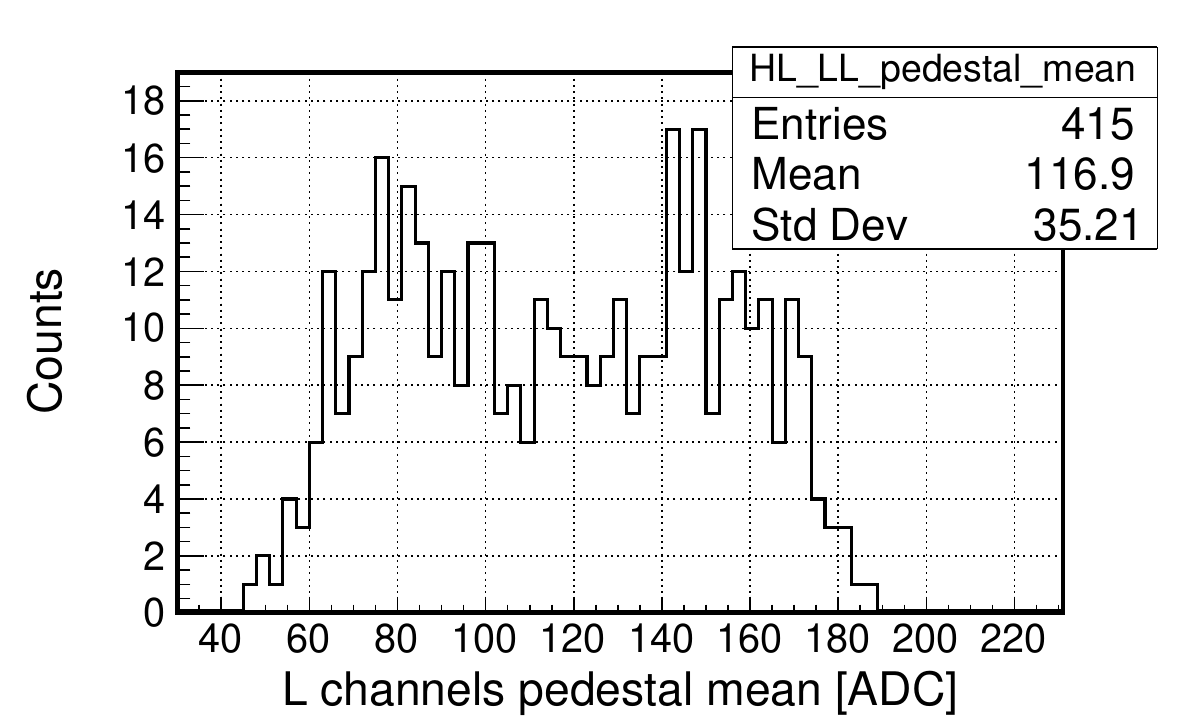}
}
\caption{(a) An example of pedestal distribution in HH12 channel Layer 7. (b) An example of pedestal distribution in HL12 channel Layer 7. (c) and (d) demonstrate statistical results of all the mean value of pedestals in high-gain channels and low-gain channels from 1 day’s test. The conversion factor in the high gain channel is 20.6 ADC/fC, while the factor in the low gain channel is 0.53 ADC/fC, according to ~\cite{bib:58}.}
\label{figure 10}
\end{figure}

The distribution of pedestal values provides insights into the electronic baseline and equivalent noise characteristics of each channel. Fig.~\ref{figure 10} illustrates two examples of pedestal distributions for a high-gain channel ~\ref{figure 10a} and a low-gain channel ~\ref{figure 10b}. The high-gain channel exhibits a mean value of approximately 1076 ADC counts with a standard deviation of 7 counts, while the low-gain channel has a mean value of about 174 ADC counts with a standard deviation of less than 1 count. These values reflect the different gain settings and noise levels of the two channel types.

Further analysis reveals significant variations in the mean pedestal values between different channels, as depicted in subsequent figures (~\ref{figure 10c} and ~\ref{figure 10d}) showing the mean values derived from Gaussian fitting of high-gain and low-gain channels during a one-day test. These variations can be attributed to several factors, including non-uniformity of temperature across the BGO crystals and APDs at different spatial locations, disparities in gain coefficients among APDs despite the common bias high voltage, and potential slight light leakage effects. Temperature variations can affect the performance of both the BGO crystals and the APDs, leading to changes in their response. Variations in the gain coefficients of the APDs can arise from manufacturing tolerances or differences in their operating conditions and recommended setup. Light leakage can introduce unwanted background signals, affecting the pedestal values. These factors exert various degrees of influence on the signal generation and amplification processes, thereby endowing each channel's pedestal with distinct characteristics.

\begin{figure}[!htb]
\subfigure[]{
\label{figure 11a}
\includegraphics[width=0.45\hsize]{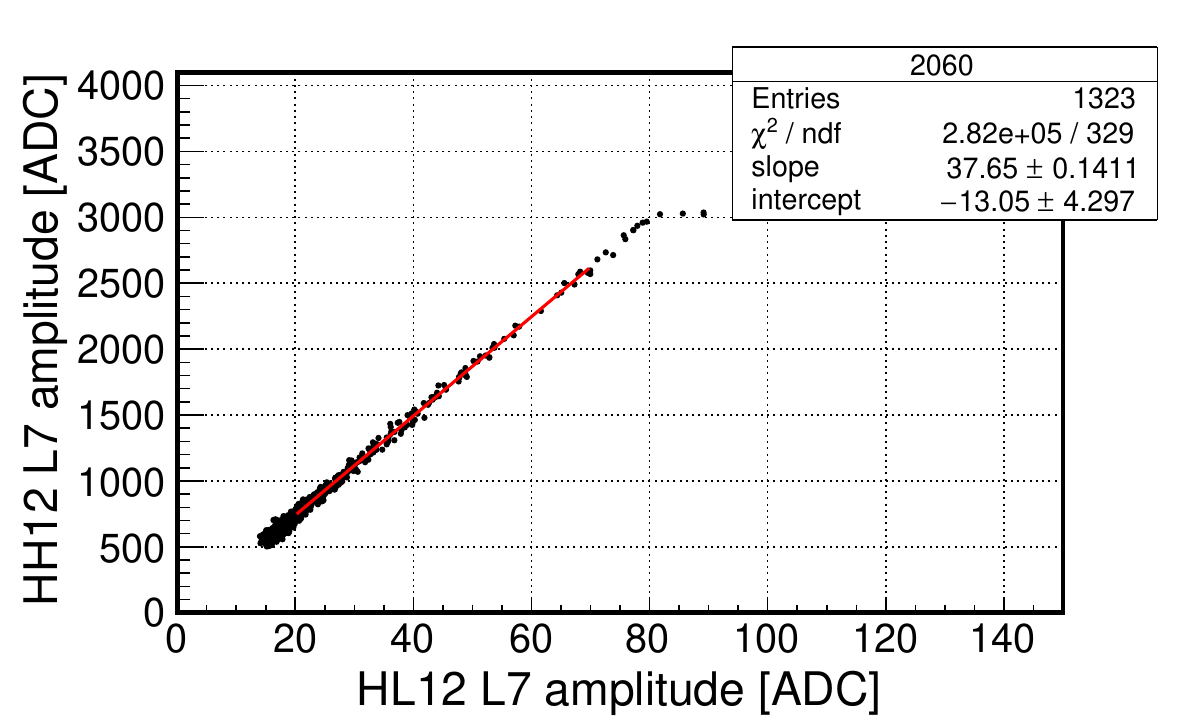}
}
\subfigure[]{
\label{figure 11b}
\includegraphics[width=0.45\hsize]{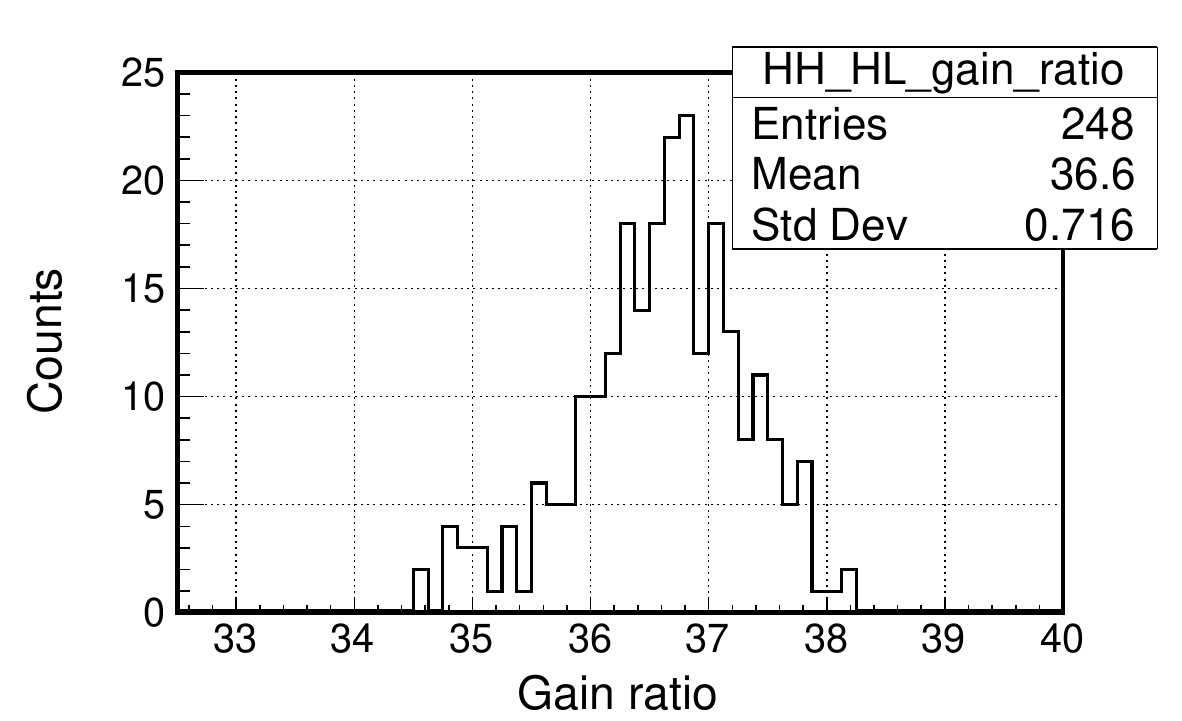}
}
\caption{(a) A typical relationship between the effective amplitudes in HL12 channel Layer 7 and HH12 channel Layer 7. (b) A statistical result of all the gain ratios of signals in high-gain channels and low-gain channels from 1 day’s test.}
\label{figure 11}
\end{figure}

To determine the gain coefficients between the high-gain and low-gain channels, a linear fitting analysis is performed. The amplitudes relationship of the two output signals from the same APD without attenuation filter coverage is presented in Fig.~\ref{figure 11a}. The linear region of the relationship between the two signals is fitted using a first-order polynomial function. Fig.~\ref{figure 11b} collects and demonstrates the slope value after linear fitting, and their average is approximately 37.5, indicating that these gain coefficients are similar and maintain reasonable consistency with the original design specifications. Given that the energy deposition of muons in cosmic rays is generally relatively small, only the uncovered APDs with high-gain can produce detectable MIP signals. Therefore, it is challenging to calibrate the remaining two coefficients (HL-LH ratio and LH-LL ratio) using a single data file.

\begin{figure}[!htb]
     \centering
     \includegraphics[width=0.5\hsize]{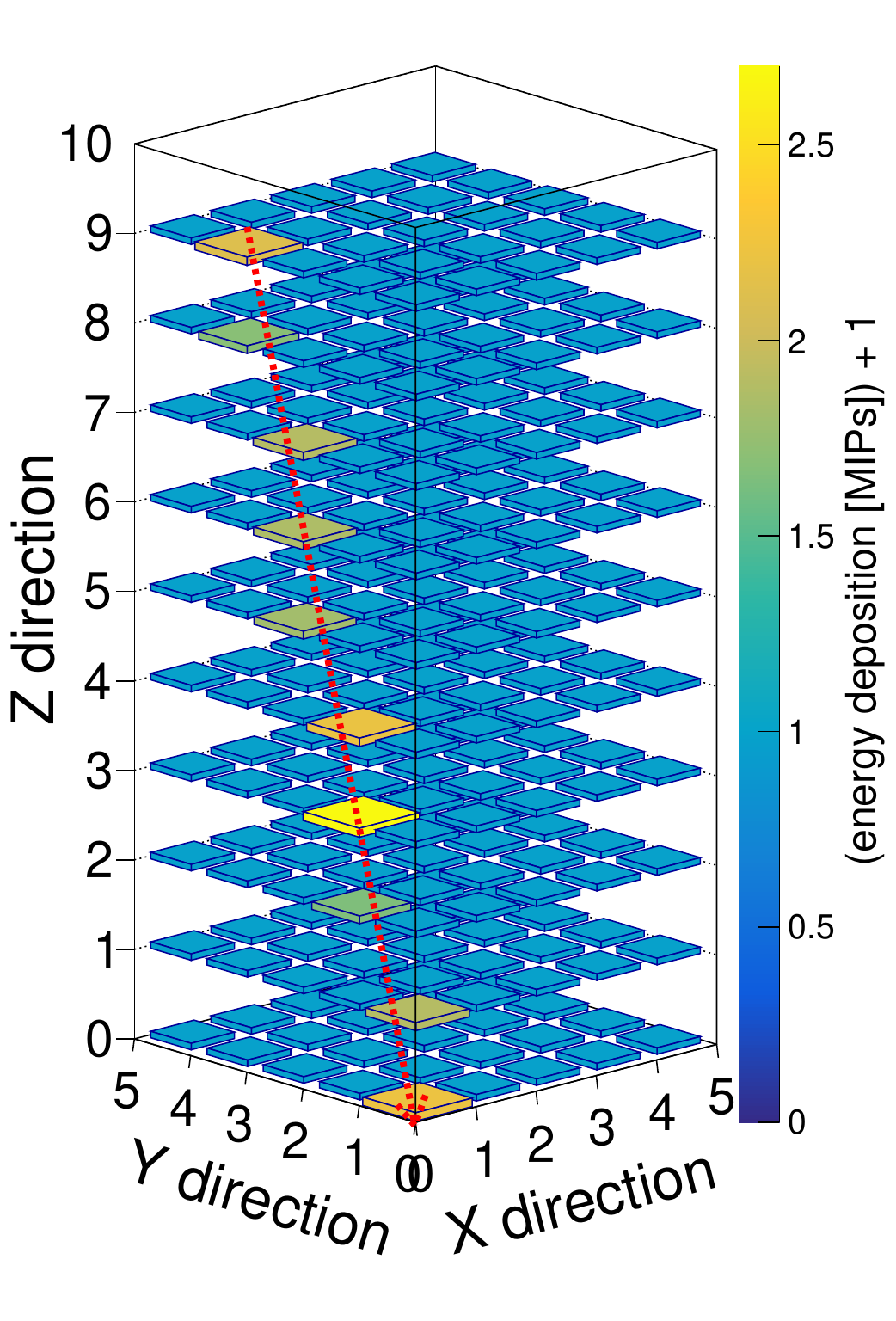}
     \caption{A typical event which is selected as a MIP incident event.}
     \label{figure 12}
\end{figure}

\begin{figure}[!htb]
\subfigure[]{
\label{figure 13a}
\includegraphics[width=0.45\hsize]{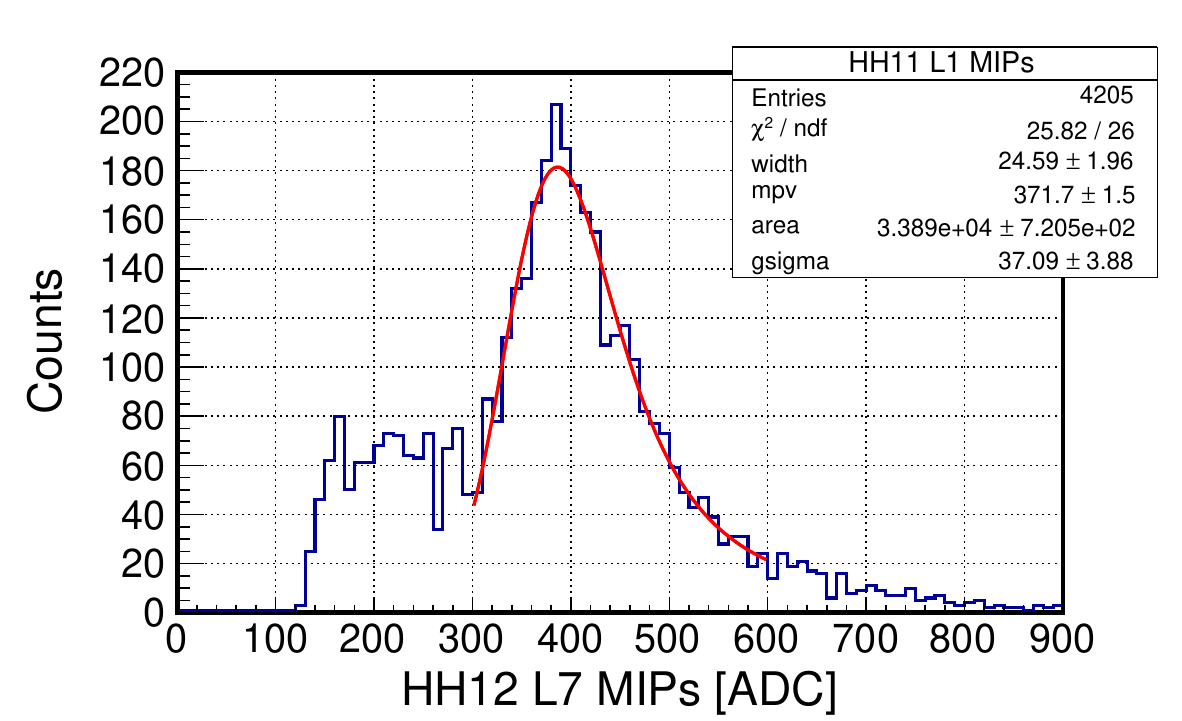}
}
\subfigure[]{
\label{figure 13b}
\includegraphics[width=0.45\hsize]{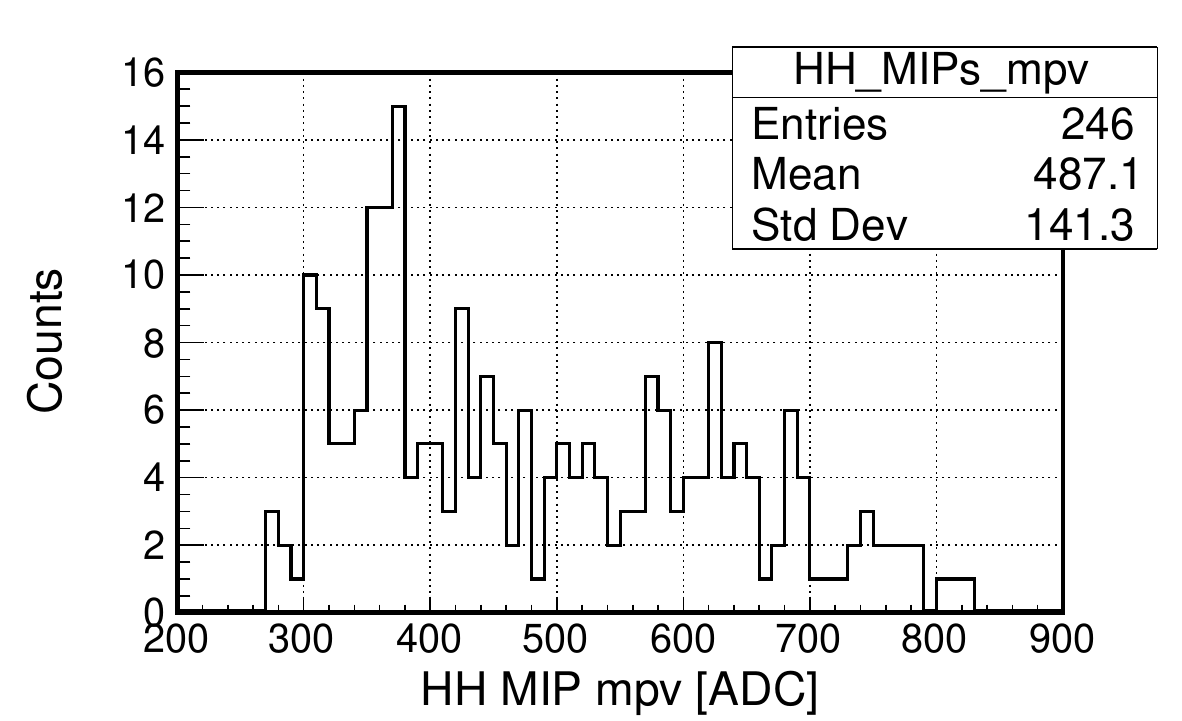}
}
\caption{(a) A typical MIPs distribution after subtracting its own pedestal in HH12 channel Layer 7. (b) A statistical result of all the most probable value of MIPs in high-gain channels from 1 day’s test. The conversion factor in the high gain channel is 20.6 ADC/fC, according to ~\cite{bib:58}.}
\label{figure 13}
\end{figure}

A group of specific selection criteria are defined to identify MIP events within the dataset. A MIP event is defined as an event in which at least 7 out of the 10 detection layers output effective signals, with no more than 3 over-threshold channels per layer, and there should be at least 2 of the first 3 layers and 2 of the last 3 layers meet these criteria. This selection ensures that the identified events correspond to muons traversing the entire calorimeter. Fig.~\ref{figure 12} depicts a representative event that satisfies these requirements. The selected MIP events are then analyzed channel by channel. The signal distribution for each channel is fitted using a Landau convoluted Gaussian function, which is commonly used to model the energy loss distribution of charged particles traversing a material. The Most Probable Value (MPV) obtained from the Landau distribution component represents its response to MIP in a given crystal. For example, the HH12L7 channel illustrated in Fig.~\ref{figure 13a} exhibits an MPV of approximately 370 ADC counts after pedestal subtraction, corresponding to an estimated input charge of about 18 fC for MIP signals.

Analysis of all the HH channels reveals that each channel has a characteristic MPV, as shown in Fig.~\ref{figure 13b}. These variations in MPV can be attributed to several factors, similar to the variations observed in the pedestal values. These factors include temperature fluctuations, variations in bias voltage between devices, and environmental stray light interference.

\begin{figure}[!htb]
\subfigure[]{
\label{figure 14a}
\includegraphics[width=0.95\hsize]{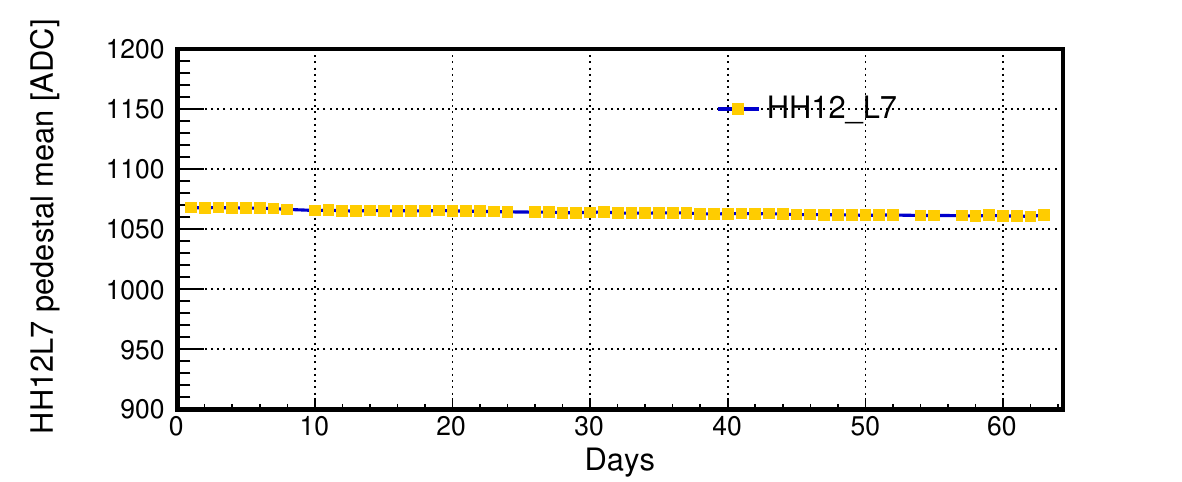}
}
\subfigure[]{
\label{figure 14b}
\includegraphics[width=0.95\hsize]{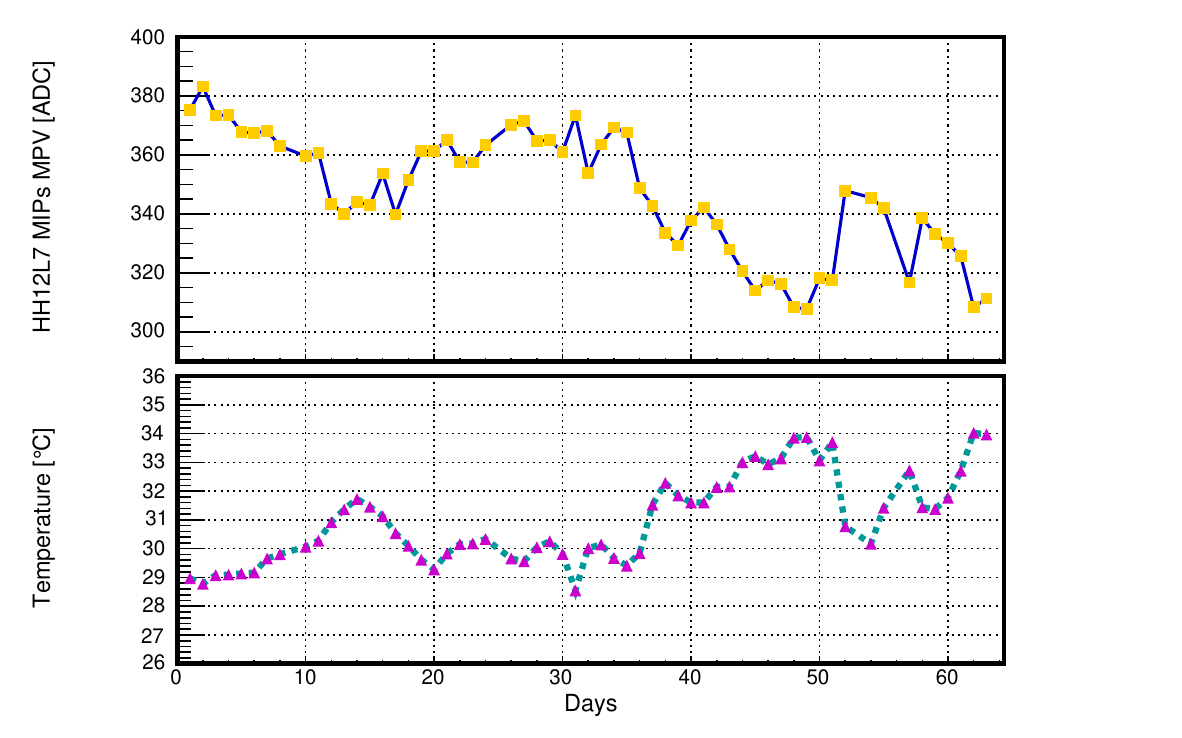}
}
\caption{(a) Variation of the mean value of pedestal in HH12 channel Layer 7 over time. (b) Variation of the MPV of MIPs and temperature (dashed lines) in HH12 channel Layer 7 over time.}
\label{figure 14}
\end{figure}

The long-term stability of the calorimeter's response was also investigated. The cosmic ray experiment has been running for two months, allowing for the observation of long-term trends. Some parameters, such as the pedestal and MPV of Landau component, showed gradually emerging regular patterns over time. Fig.~\ref{figure 14} demonstrates that the MPV of individual channels varied within a relatively small range, while the pedestal values remained remarkably stable throughout the testing period. The temperature of the channels, presented as dotted lines in Fig.~\ref{figure 14b} with reference to the right axis, exhibits a significant correlation with the observed signal variations, further supporting the hypothesis that temperature fluctuations play a significant role in the observed channel-to-channel variations. This long-term stability is crucial for ensuring the reliable performance of the calorimeter over extended periods of operation. Consequently, more stringent temperature control requirements have been put forward for the satellite's payload platform.

\section{Conclusion}\label{sec.V}

To investigate the physical characteristics of high-energy gamma-rays in cosmic environment and explore the fundamental nature of dark matter particles, we propose the development of a Very Large Area gamma-ray Space Telescope (VLAST), which will serve as China's next generation flagship satellite platform for gamma-ray space-based astronomical observation. A key component of VLAST is a high-energy imaging calorimeter, which requires high energy resolution and large dynamic range. For this purpose, a proof-of-principle prototype calorimeter has been developed along the technical approach of high granular crystal scheme. The prototype utilizes an array of 30 mm cubic BGO crystals as scintillators, coupled with a custom-designed electronics system featuring dual-APD dual-gain readout scheme for each crystal. This two APDs configuration, combined with attenuation filters, enables a wide dynamic range, crucial for detecting both low-energy and high-energy gamma rays. Initial testing of the prototype, including LED luminescence tests and ground-based cosmic ray measurements, has demonstrated promising results. The noise level of the dual-APD configuration has been determined to be approximately 0.1 MIPs, while maintaining an exceptional dynamic range of 2 × 10$\rm^6$ for the complete readout system. This wide dynamic range allows the calorimeter to detect signals ranging from the reasonable small to the much larger energy depositions of high-energy cosmic rays.

Future optimization efforts will focus on refining several key aspects of the calorimeter design and performance. (1) Precise channel-by-channel calibration of the effective sensitive regions will be implemented using a set of dedicated light intensity monitoring system. This calibration will ensure accurate energy measurements across the entire calorimeter. (2) An improved attenuator filter design will be applied to optimize the balance between a wide enough overlap region between the high-gain and low-gain channels and a large dynamic range coverage. (3) Ongoing improvements in thermal management and grounding configurations will further enhance the stability and performance of the readout electronics. Finally, a prospective beam test is planned to comprehensively evaluate the calorimeter's performance characteristics under controlled conditions with known particle beams. This beam test will provide crucial data for validating the calorimeter's design and optimizing its performance for the VLAST mission.

\end{document}